\documentclass[12pt]{colt2019} 

\title{mRSC: Multi-dimensional Robust Synthetic Control}
\usepackage{times}

\coltauthor{%
 \Name{Muhummad Amjad} \Email{mamjad@mit.edu}\\ 
 \addr  Department of Electrical Engineering and Computer Science\\
       Massachusetts Institute of Technology\\
       Cambridge, MA 02139, USA \\ \\
 \Name{Vishal Misra} \Email{vishal.misra@columbia.edu}\\
       \addr Department of Computer Science\\
       Columbia University\\
       New York, NY 10027, USA \\ \\ 
 \Name{Devavrat Shah} \Email{devavrat@mit.edu}\\
       \addr Department of Electrical Engineering and Computer Science\\
       Massachusetts Institute of Technology\\
       Cambridge, MA 02139, USA \\ \\ 
 \Name{Dennis Shen} \Email{deshen@mit.edu}\\ 
       \addr Department of Electrical Engineering and Computer Science\\
       Massachusetts Institute of Technology\\
       Cambridge, MA 02139, USA
}

\usepackage{graphicx}
\usepackage{tikz}
\usetikzlibrary{fit,positioning,arrows,automata,calc}
\tikzset{
  main/.style={circle, minimum size = 5mm, thick, draw =black!80, node distance = 10mm},
  connect/.style={-latex, thick},
  box/.style={rectangle, draw=black!100}
}

\usepackage{amsmath}
\usepackage{amsfonts}
\usepackage[colorinlistoftodos]{todonotes}
\usepackage{bm}
\usepackage{breqn}
\usepackage{physics}
\usepackage{enumitem}
\usepackage{bbold}
\usepackage{epigraph}

\usepackage{csquotes}
\usepackage{hyperref}       
\usepackage{url}            
\usepackage{booktabs}       
\usepackage{multirow}
\usepackage{amsfonts}       
\usepackage{nicefrac}       
\usepackage{microtype}      
\usepackage{color}
\usepackage{scalerel}
\usepackage[toc,page]{appendix}
\usepackage[font=small,labelfont=bf]{caption}
\usepackage{blindtext}
\usepackage{amssymb}
\usepackage{mathtools}
\usepackage{dsfont}
\allowdisplaybreaks
\usepackage{float}
\usepackage{graphicx}
\usepackage{subcaption}
\graphicspath{ {images/} }
\usepackage{letltxmacro}%
\usepackage{thmtools,thm-restate}
\usepackage{relsize}
\usepackage{algpseudocode,algorithm,algorithmicx}
\usepackage{algpseudocode}
\usepackage[ruled]{algorithm2e}

\newtheorem*{thm*}{Theorem}

\newtheorem{prop}{Proposition}[section]
\newtheorem*{prop*}{Proposition}
\newtheorem{property}{Property}[section]

\DeclareMathOperator*{\argmin}{arg\,min}

\newcommand{\Reals}{\mathbb{R}}

\newcommand{\bI}{\boldsymbol{I}}

\newcommand{\bU}{\boldsymbol{U}}
\newcommand{\bV}{\boldsymbol{V}}
\newcommand{\bW}{\boldsymbol{W}}

\newcommand{\bY}{\boldsymbol{Y}}
\newcommand{\bZ}{\boldsymbol{Z}}
\newcommand{\bM}{\boldsymbol{M}}

\newcommand{\bT}{\boldsymbol{T}}

\newcommand{\tM}{\mathcal{\bM}}

\newcommand{\tT}{\mathcal{\bT}}
\newcommand{\tZ}{\mathcal{\bZ}}

\newcommand{\hM}{\widehat{M}}

\newcommand{\bhM}{\widehat{\bM}}

\newcommand{\distas}[1]{\mathbin{\overset{#1}{\kern\z@\sim}}}%
\newsavebox{\mybox}\newsavebox{\mysim}

\newcommand{\distras}[1]{%
  \savebox{\mybox}{\hbox{\kern3pt$\scriptstyle#1$\kern3pt}}%
  \savebox{\mysim}{\hbox{$\sim$}}%
  \mathbin{\overset{#1}{\kern\z@\resizebox{\wd\mybox}{\ht\mysim}{$\sim$}}}%
}

\newcommand{\hrho}{\widehat{\rho}}

\newcommand{\hbeta}{\widehat{\beta}}

\newcommand{\Ex}{\mathbb{E}}
\newcommand{\Pb}{\mathbb{P}}

\begin{document}

\maketitle

\small{
\begin{abstract}

When evaluating the impact of a policy (e.g., gun control) on a metric of interest (e.g., crime-rate), it may not be possible or feasible to conduct
a randomized control trial. In such settings where only observational data is available, Synthetic Control (SC) methods \cite{abadie1, abadie2, abadie3} provide a popular data-driven approach to estimate a ``synthetic'' or ``virtual'' control by combining measurements of ``similar'' alternatives or units (called ``donors'').

Recently, Robust Synthetic Control (RSC) \cite{rsc1} was proposed as a
generalization of SC to overcome the challenges of missing data and
high levels of noise, while removing the reliance on expert domain
knowledge for selecting donors. However, both SC and RSC (and its
variants) suffer from poor estimation when the pre-intervention period
is too short.

As the main contribution of this work, we propose a generalization of unidimensional RSC to multi-dimensional Robust Synthetic Control, mRSC. Our proposed mechanism, mRSC, incorporates multiple types of measurements (or metrics) in addition to the measurement of interest for estimating a synthetic control, thus overcoming the challenge of poor inference due to limited amounts of pre-intervention data. We show that the mRSC algorithm, when using $K$ relevant metrics, leads to a consistent estimator of the synthetic control for the target unit of interest under any metric. 
Our finite-sample analysis suggests that the post-intervention (testing) mean-squared error (MSE) of our predictions vanishes faster than the RSC algorithm by a factor of $\sqrt{K}$.
%
Additionally, we propose a principled scheme to combine multiple metrics of interest via a diagnostic test that evaluates if adding a metric can be expected to result in improved inference.

Our mechanism for validating mRSC performance is also an important and related contribution of this work: time series prediction. We propose a method to predict the future evolution of a time series based on limited data when the notion of time is relative and not absolute, i.e., where we have access to a donor pool that has already undergone the desired future evolution. 

We conduct extensive experimentation to establish the efficacy of mRSC in three different scenarios: predicting the evolution of a metric of interest using synthetically generated data from a known factor model, and forecasting weekly sales and score trajectories of a Walmart store and Cricket game, respectively. 

\end{abstract}


}


\tableofcontents

\newpage

\maketitle

\section{Introduction} \label{sec:intro}

Quantifying the causal effect of interventions is a problem of interest across a wide array of domains. 
From policy making to engineering and medicine, estimating the effect of an intervention is critical for innovation and understanding 
existing systems. In any setting, with or without an intervention, we only observe one set of outcomes. In casual 
analysis, the fundamental problem is that of estimating what wasn't observed, i.e., the {\em counterfactual}. 
In order to estimate the counterfactual, observational studies rely on the identification (or estimation) of a {\em control} unit. This can be achieved by relying on expert domain knowledge, or via techniques such as {\em matching} the target unit to existing control units (called {\em donors}) on covariates features or propensity scores (see \cite{rubin1973, rosenbaum1983}). A popular data-driven approach to estimating the control is known as the Synthetic Control (SC) method  \cite{abadie1, abadie2, abadie3}. SC assigns convex weights to the donors such that the resulting {\em synthetic} unit most closely matches the target unit according to a chosen metric of interest. A generalization of this approach, Robust Synthetic Control (RSC) \cite{rsc1}, removes the convexity constraint and guarantees a consistent estimator that is robust to missing data and noise. 

While SC, its many variants, and RSC exhibit attractive properties, they all suffer for poor estimation when 
the amount of training data (i.e., the length of the pre-intervention period) is small (e.g., see Figure \ref{fig:walmart} in Section \ref{sec:retail}). In many of these scenarios with little pre-intervention data, there may be other data available which is related to the type of data (or metric) of interest. For example, we might be interested in crime-rate and might also have median household income and high-school graduation rate data available. Therefore, one remedy to 
the limited pre-intervention data is to utilize data from multiple metrics. 

%
%
%
%

\subsection{Contributions}

As the main contribution of this work, we propose multi-dimensional Robust Synthetic Control (mRSC), a generalization of unidimensional RSC. Unlike standard SC-like methods, mRSC incorporates multiple types of data in a principled manner to overcome the challenge of forecasting the counterfactual from limited pre-intervention data. 
We show that the mRSC method is a natural consequence of the popular factor model that is commonly utilized in the field of econometrics. Through this connection, we provide a data-driven falsifiability test for mRSC, which determines whether mRSC can be expected to produce a consistent estimation of the counterfactual evolution across all metrics for a target unit (exposed to intervention) in both the pre- and post-intervention stages (see Section \ref{sec:falsifiability}). 

Further, we demonstrate that prediction power of mRSC improves over RSC as the number of included relevant data types increases. 
Specifically, the mean-squared error (MSE) vanishes at a rate scaling with $\sqrt{K}$ for the post-intervention period, where $K$ denotes the number of metrics (see Theorem \ref{thm:post_int} in Section \ref{sec:results});
we highlight that $K=1$ reduces to the vanilla RSC framework. 
Stated another way, incorporating different types of relevant data (metrics) improves the generalization error since the number of training data increases by a factor of $K$.
Additionally, desirable properties of the RSC algorithm, namely robustness to missing data and noise, carry over to our mRSC framework. Finally, we conduct extensive experimentation to establish the efficacy of this generalized method in comparison to RSC via synthetically generated datasets and two real-world case-studies: product sales in retail and score trajectory forecasting in the game of Cricket. We now summarize these contributions in greater detail.

\medskip
\noindent {\bf Model.} We consider a natural generalization of the factor (or latent variable) model considered in the literature cf. \cite{rsc1}, where the measurement associated with a given unit at a given time for a given metric is a function of the features associated with the particular unit, time, and metric. A special case of such models, the linear factor models, are the {\em bread and butter} models within the Econometrics literature, and are assumed in \cite{abadie1, abadie2, abadie3}. We argue that as long as the function is Lipschitz and the features belong to a compact domain, then the resulting factor model can be well approximated by low-rank third-order tensor (see Proposition \ref{prop:low_rank_approx} in Section \ref{sec:model_mean_tensor_assumptions}) with the dimensions of the tensor corresponding to the unit, time, and metric, respectively. Therefore, to simplify this exposition, we focus on the setup where the underlying model is indeed a low-rank tensor. We assume that our observations are a corrupted version of the ground truth low-rank tensor with potentially missing values. More specifically, we assume we have access to pre-intervention data for all units and metrics, but only observe the post-intervention data for the donor units. Our goal is to estimate the ``synthetic control'' using all donor units and metrics in the pre-intervention period. The estimated synthetic control can then help estimate the {\em future} (post-intervention) ground-truth measurements for all metrics associated with the treatment unit. 

\medskip
\noindent {\bf Algorithm and analysis.} The current SC literature \cite{abadie1, abadie2, abadie3, rsc1} relies on the following 
key assumption: for any metric of interest, the treatment unit's measurements across time can be expressed as a linear combination of the donor units' measurements. In this work, we start by arguing that we need not make this assumption as it holds naturally under the low-rank tensor model (Proposition \ref{prop:linear_comb} in Section \ref{sec:results}).  Furthermore, the low-rank tensor model suggests that the identical linear relationship holds across all metrics simultaneously with high probability. That is, in order to utilize measurements from other metrics to estimate the synthetic control for a given metric and unit of interest, we can effectively treat the measurements for other metric as additional measurements for our metric of interest! In other words, the number of pre-intervention measurements is essentially multiplied by the number of available metrics. The resulting mRSC algorithm is natural extension of RSC but with multiple measurements. Using a recently developed method for analyzing error-in-variable regression in the high-dimensional regime, we derive finite-sample guarantees on the performance of the mRSC algorithm. 
We conclude that the post-intervention MSE for mRSC decays faster than that of RSC by a factor of $\sqrt{K}$, where $K$ is the number of available metrics (see Theorem \ref{thm:post_int} in Section \ref{sec:results}).
This formalizes the intuition that data from other metrics can be treated as belonging to the original metric of interest. In summary, mRSC provides a way to overcome limited pre-intervention data by simply utilizing pre-intervention data across other metrics.

\medskip
\noindent {\bf Experiments: synthetic data.} To begin, we utilize synthetically generated data as per our factor model to both verify the tightness of our theoretical results and utility of our diagnostic test, which evaluates when mRSC is applicable. We argue that flattening the third-order tensor of measurement data into a matrix (by stacking the unit by time slices across metrics) is valid if the resulting matrix exhibits a similar singular value spectrum (rank) as that of the matrix with only a single metric of interest (see Section \ref{sec:falsifiability}). Our results demonstrate that data generated as per our model pass the diagnostic test; the test, however, fails when the data does not come from our model. Finally, our empirical findings support our theoretical results. 

%

\medskip
\noindent {\bf Experiments: retail.} Next, we study the efficacy of mRSC in a real-world case-study regarding weekly sales data across multiple departments at various Walmart stores (data obtained from Kaggle \cite{kaggle-walmart}). The store locations represent units, weeks represent time, and product-departments represent different metrics. Our goal is to forecast the sales of a given product-department at a given location using a subset of the weekly sales reports as the pre-intervention period. 
We study the performance of mRSC and compare it with RSC across different intervention times and different department-product subgroups, which represent the available metrics. Across all our experiments, we consistently find that mRSC significantly outperforms RSC when the pre-intervention data is small; however, the two methods perform comparably in the presence of substantial pre-intervention data. These empirical findings are in line with our theoretical results, i.e., in the presence of sparse training data, mRSC provides significant gains over RSC by utilizing information from auxiliary metrics. Specifically, Table \ref{table:avg_mse} demonstrates that the test prediction error of mRSC is 5-7x better compared to vanilla RSC when very limited pre-intervention data is available. Further, even if the pre-intervention data is significant, mRSC's test prediction error continues to outperform that of RSC. 

\medskip
\noindent {\bf Experiments: cricket.} We consider the task of forecasting score trajectories in cricket. Since cricket has {\em two} metrics of interest, runs scored and wickets lost, cricket is an ideal setting to employ mRSC. 
We first provide a brief primer on cricket as we do not expect the reader(s) to be familiar with the game. This is followed by an explanation as to why forecasting the score trajectory is a perfect fit for mRSC. We conclude with a summary of our findings. 

\medskip
\noindent {\em Cricket 101.} Cricket, as per a Google Search on ``how many fans world-wide watch cricket", is the second most popular sport
after soccer. It is played between two teams and is inherently asymmetric 
in that both teams take turns to bat and bowl. Among the multiple formats with varying durations of a game, we focus on one format
called ``50 overs Limited Over International (LOI)''. Here, each team bats for an ``inning'' of 50 overs or 300 balls, during 
which, at any given time, two of its players (batsmen) are batting to score runs. Meanwhile, the other team fields all of its 11 players,
one of whom is bowling and the others are fielding with the goal of getting the batsmen out or preventing the batsmen from scoring. Each batsman can get out at most once, and an inning ends when either 300 balls are bowled or 10 batsmen are out. At the end, the team that scores more runs wins. For more details relevant to this work, refer to Section \ref{sec:cricket}.

\medskip
\noindent{\em Forecasting trajectory using mRSC.} As an important contribution, we utilize the mRSC algorithm to forecast an entire score trajectory of a partially observed cricket inning. As the reader will notice, the utilization of mRSC for forecasting the trajectory of game is generic, and is likely to be applicable to other games such as basketball, which would be an interesting future direction. To that end, we start by describing how mRSC can be utilized for forecasting the score trajectory of an ``inning''. We view each inning as a unit, the balls as time, and the runs scored and wickets lost as two different metrics of interest. Consequently, past innings are effectively donor units. The inning of interest, for which we might have observed scores/wickets up to some number of balls (or time), is the treatment unit, and the act of forecasting is simply coming up with a ``synthetic control'' for this unit in order to estimate the score/wicket trajectory for remainder of the game. 

We collect data for historical LOI over the period of 1999-2017 for over 4700 innings. Each inning can go to 300 balls long for both metrics: runs and wickets. We find that the approximately low-rank structure of the matrix with only scores (4700 by 300) and that with both scores and wickets (4700 by 600) remains the same (see Figure \ref{fig:singvals}). This suggests that mRSC can be applicable for this scenario.

Next, we evaluate the predictive performance on more than 750 recent (2010 to 2017) LOI innings. For consistency of comparisons, we conduct a detailed study for forecasting at the 30th over or 180 balls (i.e., 60\% of the inning). We measure performance 
through the mean-absolute-percentage-error (MAPE) and R-square ($R^2$). The median of the MAPE of our forecasts varies 
from $0.027$ or $2.7$\% for 5 overs ahead to $0.051$ or $5.1$\% for 20 overs ahead. While this is significant in its own right, 
to get a sense of the {\em absolute} performance, we evaluate the $R^2$ of our forecasts. We find that the $R^2$ of 
our forecast varies from $0.92$ for 5 overs to $0.79$ for 15 overs. That is, for the 5 over forecast, we can explain $92$\% of 
``variance'' in the data, which is surprisingly accurate. Further, we establish the value of the de-noising and regression steps 
in the mRSC algorithm. Through one important experiment, we demonstrate that a commonly held belief---i.e., the donor pool should only comprise of innings involving the team we are interested in forecasting for---actually
leads to worse estimation.  
This is related to {\em Stein's Paradox} \cite{stein1}, which was first observed in baseball. Finally, using examples of actual 
cricket games, we show that the mRSC algorithm successfully captures and reacts to nuances of cricket. 

\subsection{Related Works} \label{sec:related}

There are three bodies of literature relevant to this work: matrix estimation for sequential data inference, synthetic control, and 
multi-dimensional causal inference.

In this work, we rely on low-rank estimation which is a key subproblem in the rich domain of matrix estimation. We refer the interested reader to \cite{lee2017, amjad2018, usvt, BorgsChayesLeeShah17} for extensive surveys of the matrix estimation literature. These works, in particular \cite{lee2017, amjad2018} also use the latent variable model (LVM) which is at the heart of the mRSC model presented in this work. In the context of sequential data inference and forecasting, matrix estimation techniques have recently been proposed as important subroutines. Using low-rank factorizations to learn temporal relationships has been addressed in \cite{dhillon2016, xie16missingtimeseries, rallapalli2010, chen2005}. However, these works make specific assumptions about the particular time series under consideration which is a major difference to our work where we make no assumptions about the specifics of the temporal evolution of the data. \cite{timeseriesMatrixEst, amjad2018} present algorithms closely related to the mRSC algorithm presented in this work to capture relationships across both time- and location without making assumptions about the particular form of temporal relationships. However, in \cite{timeseriesMatrixEst} the authors consider a single (one-dimensional) time series and convert it to a Page Matrix and then apply matrix estimation. The goal in that work is to estimate the future {\em only} using data from the single time series and up to present time. In this work, the notion of time is relative, i.e. there is a temporal ordering but the temporal axis is not absolute, there are many other units (donors) and the estimation of the counterfactual happens longitudinally, i.e. as a function of all other units (donors). Finally, we note that in this work the donor units have all their \emph{future} observations realized while that is certainly not the case in time series forecasting.

The data driven method to estimate ``synthetic control'' was original proposed in \cite{abadie1, abadie2} (SCM). The robust synthetic control (RSC) algorithm was presented as a recent generalization of the SCM making it robust to missing data and noise. For a detailed theoretical analysis of the RSC and an overview of the more recent developments of the synthetic control method, we refer the reader to \cite{rsc1}. A key limitation of the RSC and other SCM variants is that they are only concerned with a single metric of interest {red} ($K = 1$) and perform poorly when the pre-intervention (training) period is short. The mRSC algorithm presented in this work generalizes the RSC and allows a principled way to handle multiple related metrics of interest ($K > 1$). 

Causal inference has long been an interest for researchers from a wide array of communities, ranging from economics to machine learning (see \cite{pearl2009, rubin1974, rubin1973, rosenbaum1983} and the references therein). Recently there has been work in multidimensional causal inference, specifically building causal models that fuse multiple datasets collected under heterogeneous conditions. A general, non-parametric approach to fusing multiple datasets that handles biases has been presented in \cite{Bareinboim7345}. This body of work is similar in spirit to the contribution of our paper, namely generalizing robust synthetic control to utilize multiple metrics of interest.

Forecasting the scores and, more specifically, the winner of a cricket game has also been of interest. Works such as \cite{Bailey2006} use multi-regression models to predict the winner of a cricket game. However, predicting the winner is not a goal of this work. Our work focuses on forecasting the future average trajectory of an innings 
given its partial observation. Some prior efforts in this domain include the WASP method \cite{wasp1}, which calculates the mean future scores via 
a dynamic programming model instead of learning these solely from past data like we do. Recent efforts, such as \cite{cricviz}, have also focused on 
forecasting trajectories of the remainder of the innings using complex models that quantify specific features of an innings/game that affect scoring and 
game outcomes. However, like WASP, these recent methods are also parametric, and their accuracy is a strict function of whether the model is accurately able to capture the complex and evolving nature of the game. In summary, all known prior works on score forecasting make parametric assumptions to develop their methods. In contrast, we make minimal modeling assumptions and instead rely on the generic {\em latent variable model} to use the mRSC algorithm to capture nuances of the game of cricket.
%
%
\subsection{Organization} The rest of this work is organized as follows: we review the relevant bodies of literature next (Section \ref{sec:related}) followed by a
detailed overview of our proposed model and problem statement (Section \ref{sec:model}). Next, in Section \ref{sec:algorithm}, we describe the mRSC algorithm, which is followed by a simple
diagnostic test to determine the model applicability to a problem. We present the main theoretical results in Section \ref{sec:results} followed by a synthetic experiment and ``real-world'' retail case study to
compare the mRSC algorithm to the RSC (Section \ref{sec:experiments}). Finally, in Section \ref{sec:cricket}, we have the complete case study for the game of cricket to demonstrate the versatility of the proposed
model and the mRSC's superior predictive performance.


\section{Problem Setup} \label{sec:model}

Suppose there are $N$ units indexed by $i \in [N]$, across $T$ time periods indexed by $j \in [T]$, and $K$ metrics 
of interest indexed by $k \in [K]$. Let $M_{ijk}$ denote the ground-truth measurement of interest and $X_{ijk}$ the noisy
observation of $M_{ijk}$. Without loss of generality, let us assume that our interest
is in the measurement associated with unit $i=1$ and metric $k = 1$. Let $1 \leq T_0 < T$ represent the time instance in which unit one 
experiences an {\em intervention}. Our interest is to estimate the measurement evolution of metric one for unit one
if {\em no intervention} occurred. To do so, we utilize the measurements associated with the
``donor'' units ($2 \le i \le N$), and possibly all metrics $k \in [K]$. 

\subsection{Model description} 
For all $2 \le i \le N$, $j \in [T]$, and $k \in [K]$, we posit that
\begin{align} \label{eq:donor_model}
	X_{ijk} &= M_{ijk} + \epsilon_{ijk},
\end{align}
where $\epsilon_{ijk}$ denotes the observation noise. 
We assume that unit one obeys the same model relationship during the pre-intervention period across all metrics, i.e., for all 
$j \in [T_0]$ and $k \in [K]$, 
\begin{align} \label{eq:unit_one_model}
	X_{1jk} = M_{1jk} + \epsilon_{1jk}. 
\end{align}
If unit one was never exposed to the treatment, then the relationship described by \eqref{eq:unit_one_model} 
would continue to hold during the post-intervention period, i.e., for $j \in \{T_0+1,\dots, T\}$. 



\subsection{Structural assumptions on mean tensor} \label{sec:model_mean_tensor_assumptions}

Let $\tM = [M_{ijk}] \in \Reals^{N \times T \times K}$ be a third-order tensor denoting the underlying, deterministic means in the absence of any treatment. We assume that $\tM$ satisfies the following low-rank and boundedness properties:
\begin{property} \label{property:low_rank}
	Let $\tM$ be a third-order tensor with rank $r$, i.e., $r$ is the smallest integer such that the entries of $\tM$ can be expressed as
	\begin{align} \label{eq:low_rank_tensor}	
		M_{ijk} &= \sum_{z = 1}^r U_{i z} V_{j z} W_{k z},
	\end{align}
	where $\bU = [U_{ij}] \in \Reals^{N \times r}$, $\bV = [V_{ij}] \in \Reals^{T \times r}$, $\bW = [W_{ij}] \in \Reals^{K \times r}$. 
\end{property}

\begin{property} \label{property:boundedness}
	There exists an absolute constant $\Gamma \ge 0$ such that $| M_{ijk} | \le \Gamma$ for all $(i,j,k) \in [N] \times [T] \times [K]$. 
\end{property}

\smallskip
\noindent {\em Why $\tM$ should be low-rank.} A natural generalization of the typical {\em factor model}, which is commonly utilized in the
Econometrics literature, is the generic latent variable model (LVM), which posits that 
\begin{align} \label{eq:lvm}
M_{ijk} &= f(\theta_i, \rho_j, \omega_k).
\end{align}
Here, $\theta_i \in \Reals^{d_1}, \rho_j \in \Reals^{d_2}$, and $\omega_k \in \Reals^{d_3}$ are latent feature vectors capturing unit, time, and metric specific information, respectively, for some $d_1, d_2, d_3 \ge 1$; and the latent function $f: \Reals^{d_1} \times \Reals^{d_2} \times \Reals^{d_3 }  \to \Reals$ captures the model relationship. If $f$ is ``well-behaved'' and the latent spaces are compact, then it can be seen that $\tM$ is approximately low-rank. This is made more rigorous by the following proposition. 

\begin{prop} \label{prop:low_rank_approx}
	Let $\tM$ satisfy \eqref{eq:lvm}. Let $f$ be an $\mathcal{L}$-Lipschitz function with $\theta_i \in [0,1]^{d_1}, \rho_j \in [0,1]^{d_2}$, and $ \omega_k \in [0,1]^{d_3}$ for all $(i,j,k) \in [N] \times [T] \times [K]$. Then, for any $\delta > 0$, there exists a low-rank third-order tensor $\tT$ of rank 
$r \le C  \cdot \delta^{-(d_1 + d_3)}$ such that
\begin{align*}
	\norm{\tM - \tT}_{\emph{max}} &\le 2 \mathcal{L}  \delta. 
\end{align*}
Here, $C$ is a constant that depends on the latent spaces $[0,1]^{d_1}$ and $[0,1]^{d_3}$, ambient dimensions $d_1$ and $d_3$, and Lipschitz constant $\mathcal{L}$.
\end{prop}
In Section \ref{sec:results}, we will demonstrate that the low-rank property of $\tM$ is central to establishing that $\tM$ satisfies the key property of synthetic control-like settings, i.e., the target unit can be expressed as a linear combination of the donor pool across \textit{all metrics} (see Proposition \ref{prop:linear_comb}). Hence, in effect, observations generated via a generic latent variable model, which encompass a large class of models, naturally fit within our multi-dimensional synthetic control framework.


\subsection{Structural assumptions on noise}

Before we state the properties assumed on $\epsilon = [\epsilon_{ijk}] \in \Reals^{N \times T \times K}$, we first define an important class of random variables/vectors. 

\begin{definition}
For any $\alpha \geq 1$, we define the $\psi_{\alpha}$-norm of a random variable $X$ as 
\begin{align} \label{eq:alpha_norm}
	\norm{X}_{\psi_{\alpha}} &= \inf \Big \{ t > 0: \Ex \exp(|X|^{\alpha} /t^{\alpha}) \le 2 \Big \}.
\end{align}
If $\norm{X}_{\psi_{\alpha}} < \infty$, we call $X$ a $\psi_{\alpha}$-random variable. 
More generally, we say $X$ in $\mathbb{R}^n$ is a $\psi_{\alpha}$-random vector if the one-dimensional marginals $\langle X, v \rangle$ are $\psi_{\alpha}$-random variables for all fixed vectors $v \in \mathbb{R}^n$. 
We define the $\psi_{\alpha}$-norm of the random vector $X \in \mathbb{R}^n$ as
\begin{align} \label{eq:alpha_vector_norm}
	\norm{X}_{\psi_{\alpha}} &= \sup_{v \in \mathcal{S}^{n-1}} \norm{ \langle X, v \rangle }_{\psi_{\alpha}},
\end{align}
where $\mathcal{S}^{n-1} := \{ v \in \mathbb{R}^n: \norm{v}_2 = 1\}$ denotes the unit sphere in $\mathbb{R}^n$ and $\langle \cdot, \cdot \rangle$ denotes the inner product. 
Note that $\alpha = 2$ and $\alpha =1 $ represent the class of sub-gaussian and sub-exponential random variables/vectors, respectively. 
\end{definition}
We now impose the following structural assumptions on $\epsilon$. For notational convenience, we will denote $\epsilon_{\cdot, j, k} \in \Reals^N$ as the column fiber of $\epsilon$ at time $j$ and for metric $k$. 

\begin{property} \label{property:noise.1}
	Let $\epsilon$ be a third-order tensor where its entries $\epsilon_{ijk}$ are mean-zero, $\psi_{\alpha}$-random variables  (for some $\alpha \ge 1$) with variance $\sigma^2$, that are independent across time $j \in [T]$ and metrics $k \in [K]$, i.e., there exists an $\alpha \ge 1$ and $K_{\alpha} < \infty$ such that $\| \epsilon_{\cdot, j, k} \|_{\psi_{\alpha}} \le K_{\alpha}$ for all $j \in [T]$ and $k \in [K]$. 
\end{property}

\begin{property} \label{property:noise.2}
	For all $j \in [T]$ and $k \in [K]$, let $\| \Ex [ \epsilon_{\cdot, j, k} \epsilon^T_{\cdot, j, k}] \| \le \gamma^2$.
\end{property}

\subsection{Missing data}

In addition to noise perturbations, we allow for missing data within our donor pool of observations. In particular, we observe (a possibly sparse) donor pool tensor $\tZ = [Z_{ijk}] \in \Reals^{(N-1) \times T \times K}$ where each entry $Z_{ijk}$ is observed with some probability $\rho \in (0,1]$, independent of other entries. This is made formal by the following property.

\begin{property} \label{property:masking_noise} 
	For all $(i,j,k) \in [N-1] \times [T] \times [K]$, 
	\begin{align*}
		Z_{ijk} &= \begin{cases}
      			X_{(i+1)jk} & \text{w.p. } \rho \\
      			\star & \text{w.p. } 1 - \rho
    		\end{cases}
	\end{align*}
	is sampled independently. Here, $\star$ denotes an unknown value. 
\end{property}


\subsection{Problem statement}

In summary, we observe $\tZ$, which represents the observations associated with the donor pool across the entire time horizon and for all metrics. However, we only observe the pre-intervention observations for unit one, i.e., $X_{1jk}$ for all $j \in [T_0]$ and $k \in [K]$. In order to investigate the effects of the treatment on unit one, we aim to construct a ``synthetic'' unit one to compute its counterfactual sequence of observations $M_{1jk}$ for all $j \in [T]$ and $k \in [K]$, with particular emphasis the post-intervention period (namely, $T_0 < j \le T$), using only the observations described above. We will evaluate our algorithm based on its prediction error. More specifically, we assess the quality of our estimate $\hM^{(k)}_1 \in \Reals^{T}$ for any metric $k \in [K]$ in terms of its (1) pre-intervention mean-squared error (MSE)
\begin{align} \label{eq:pre_int_mse}
	\text{MSE}_{T_0}(\hM^{(k)}_1) &= \frac{1}{T_0} \Ex \left[ \sum_{t =1}^{T_0} \left(\hM^{(k)}_{1t} - M_{1tk} \right)^2 \right],
\end{align} 
and (2) post-intervention MSE 
\begin{align} \label{eq:post_int_mse}
	\text{MSE}_{T - T_0}(\hM^{(k)}_1) &= \frac{1}{T-T_0} \Ex \left[ \sum_{t = T_0+1}^T \left(\hM^{(k)}_{1t} - M_{1tk} \right)^2 \right]. 
\end{align} 
We summarize our model assumptions in Table \ref{table:model_assumptions}\footnote{With regards to Property \ref{property:masking_noise}, we specifically mean $Z_{ijk} = X_{ijk} \cdot \mathds{1}(\pi_{ijk} =1) + \star \cdot \mathds{1}(\pi_{ijk} =0)$.}.

\begin{table} 
\caption{Summary of Model Assumptions}
\label{table:model_assumptions}
\centering
\begin{tabular}{ c c c c c }
\toprule
 \multicolumn{2}{c}{Mean Tensor $\tT$}	&	 \multicolumn{2}{c}{Observation Noise $\epsilon$} &	 \multirow{2}{*}{Masking $\tZ$}	 \\
 \cmidrule{1-2}	  \cmidrule{3-4}	
 	Low-rank & 	Boundedness 	&	$\psi_{\alpha}$-norm &  Covariance   &   \\
 \midrule
  $\text{rank}(\tM) = r$	&	$\left| M_{ijk} \right| \leq \Gamma$  	&   $\norm{\epsilon_{\cdot, j, k}}_{\psi_{\alpha}} \le K_{\alpha}$
  & $\big\| \Ex [\epsilon_{\cdot, j, k} \epsilon^T_{\cdot, j, k}] \big\| \leq \gamma^2$ 
&$\pi_{ijk} \sim \text{Bernoulli}(\rho)$\\
 Property \ref{property:low_rank}
& Property \ref{property:boundedness}
& Property \ref{property:noise.1}
& Property \ref{property:noise.2}
& Property \ref{property:masking_noise} 
\\
 \bottomrule 
\end{tabular}
\end{table}

\vspace{10pt}
\noindent {\bf Notation.}
For any general $n_1 \times n_2 \times n_3$ real-valued third-order tensor $\tT$, we denote $\tT_{\cdot, j, k}, \tT_{i, \cdot, k},$ and $\tT_{i, j, \cdot}$ as the column, row, and tube fibers of $\tT$, respectively. Similarly, we denote $\tT_{i, \cdot, \cdot}, \tT_{\cdot, j, \cdot}$, and $\tT_{\cdot, \cdot, k}$ as the horizontal, lateral, and frontal slices of $\tT$, respectively. We will denote the $n_1 \times n_2 \cdot n_3$ matrix $\bT$ as the flattened version of $\tT$, i.e., $\bT$ is formed by concatenating the $n_3$ frontal slices $\tT_{\cdot, \cdot, k}$ of $\tT$. We define $\bT_{i, \cdot}$ and $\bT_{\cdot, j}$ as the $i$-th row and $j$-th column of $\bT$, respectively. Finally, we denote $\text{poly}(\alpha_1, \dots, \alpha_n)$ as a function that scales at most polynomially in its arguments $\alpha_1, \dots, \alpha_n$. 


%
%

\section{Algorithm} \label{sec:algorithm}

\subsection{Setup}

Suppose there are $K$ datasets corresponding to $K$ metrics of interest. For each $k \in [K]$, let $\bZ^{(k)} \in \Reals^{(N-1) \times T}$ denote the $k$-th donor matrix, which contains information on the $k$-th metric for the entire donor pool and across the entire time horizon. We denote $X_1^{(k)} \in \Reals^{1 \times T_0}$ as the vector containing information on the $k$-th metric for the treatment unit (unit one), but only during the pre-intervention stage.  \\

The mRSC algorithm, to be introduced in Section \ref{sec:robust_algo}, is a generalization of the 
RSC algorithm introduced in \cite{rsc1}. The RSC algorithm utilizes a hyper-parameter $\lambda \geq 0$, 
which thresholds the spectrum of the observation matrix; this value effectively serves as a knob to trade-off between the bias 
and variance of the estimator \cite{rsc1}. The mRSC also utilizes such a hyper-parameter. In addition, it utilizes
\begin{align}
	\Delta = \text{diag}\Bigg(\underbrace{\frac{1}{\delta_1}, \dots, \frac{1}{\delta_1}}_{T_0}, \dots, \underbrace{\frac{1}{\delta_K}, \dots, \frac{1}{\delta_K}}_{T_0} \Bigg) \in \Reals^{KT_0 \times KT_0},
\end{align}
which weighs the importance of the $K$ different metrics by simply multiplying the corresponding observations. For instance, the choice of
$\delta_1 = \dots = \delta_K = 1$ renders all metrics to be equally important. On the other hand, if the only metric of interest is $k = 1$, then
setting $\delta_1 = 1$ and $\delta_2 = \dots = \delta_K = 0$ effectively reduces the mRSC algorithm to the RSC algorithm for metric one. The choice of hyper-parameters can be chosen in a data-driven manner via standard machine learning techniques such as cross-validation.

\subsection{Robust multi-metric algorithm.} \label{sec:robust_algo}

Given $\bZ^{(k)}$ and $X_1^{(k)}$ for all $k \in [K]$, we are now ready to describe the mRSC algorithm. Recall that the goal is to estimate the post-intervention observations (in the absence of any treatment) for a particular unit of interest (unit one). This estimated forecast is denoted by $\hM_1^{(k)}$ for all metrics $k \in [K]$. 

\begin{algorithm} 
\caption{Multi Robust Synthetic Control (mRSC)} \label{algorithm:mrsc}
\noindent \\

{\bf Step 1. Concatenation.} 
\begin{enumerate}

	\item Construct $\bZ \in \Reals^{(N-1)\times KT}$ as the concatenation of $\bZ^{(k)}$ for all $k \in [K]$.  
	
	\item Construct $X_1 \in \Reals^{1 \times KT_0}$ as the concatenation of $X_1^{(k)}$ for all $k \in [K]$. 

\end{enumerate}

\noindent \\
{\bf Step 2. De-noising.} 
\begin{enumerate}
	\item Compute the singular value decomposition (SVD) of $\bZ$:
	\begin{align}
		\bZ &= \sum_{i=1}^{N-1 \wedge KT} s_i u_i v_i^T. 
	\end{align}
	
	\item Let $S = \{ i : s_i \geq \lambda\}$ be the set of singular values above the threshold $\lambda$.
	
	\item Apply hard singular value thresholding to obtain
	\begin{align} \label{eq:hsvt}
		\bhM & = \frac{1}{\hrho} \sum_{i \in S} s_i u_i v_i^T. 
	\end{align}
	Here, $\hrho $ denotes the proportion of observed entries
	in $\bZ$. Further, we partition $\bhM = [\bhM^{(1)}, \dots, \bhM^{(K)}]$ into $K$ blocks of dimension $(N-1) \times T$. 
\end{enumerate}

\noindent \\
{\bf Step 3. Weighted Least Squares.} 

\begin{enumerate}
	\item Construct $\bhM_{T_0} \in \Reals^{(N-1)\times KT_0}$ as the concatenation of $\bhM^{(k)}_{\cdot, j}$ for $j \in [T_0]$ and $k \in [K]$. 
	
	\item Weight the donor matrix and treatment vector by $\Delta$, i.e.,
	\begin{align}
		\bhM_{T_0} &:= \bhM_{T_0} \cdot \Delta
		\\ X_1 &:= X_1 \cdot \Delta. 
	\end{align}
	
	\item Perform linear regression:
	\begin{align} \label{eq:linear_regression}
		\hbeta &\in \argmin_{v \in \mathbb{R}^{(N-1)}} \norm{ X_1 - v^T \bhM_{T_0}}_2^2.
	\end{align}
	
	\item For every $k \in [K]$, define the corresponding estimated (counterfactual) means for the treatment unit as 
	\begin{align}
		\hM_1^{(k)} &= \hbeta^T \bhM^{(k)}. 
	\end{align}
\end{enumerate} 
\end{algorithm}

\begin{remark} We note that all the features of the vanilla RSC algorithm naturally extend to the mRSC algorithm (see Section 3.4 of \cite{rsc1}). Specifically, properties of RSC regarding solution interpretability, scalability, and independence from using covariate information are also features of the mRSC algorithm. Effectively, the mRSC algorithm is a generalization of the RSC algorithm in that it incorporates learning based on multiple metrics of interest. 
\end{remark}

\subsection{Diagnostic: rank preservation} \label{sec:falsifiability}

While it may be tempting to include multiple metrics in any analysis, there must be some relationship between the metrics to allow for improved inference. In order to determine whether the additional metrics should be incorporated in the mRSC model, we propose a rank-preservation diagnostic. As discussed in Proposition \ref{prop:low_rank_approx}, the crucial assumption is that the mean data tensor is low-rank. For the mRSC algorithm to make sense, the row and column relationships must extend similarly across all metrics. Specifically, under a LVM, the latent row and column parameters, $\theta_i \in \mathbb{R}^{d_1}$ and $\rho_j \in \mathbb{R}^{d_2}$ (as described in Section \ref{sec:model_mean_tensor_assumptions}), must be the same for all metrics being considered. This implies that the rank of the matrices for each metric (frontal slices of the tensor) and their concatenation are identical. If, however, the row and column parameters vary with the metric, then concatenated matrix rank may increase. 

\smallskip
For concreteness, we present an analysis from a simple idealized experiment (detailed in Section \ref{sec:mrsc-synthetic}) with two metrics (metric1 and metric2) and matrix dimensions $N = 100, T = 120$. We compare the percentages of the power of the cumulative spectrum contained in the top few singular values for metric1, metric2, and two of their combinations: one where the row and column parameters are held constant and another where the parameters are different. Table \ref{table:diagnostic} shows the resulting ranks of the mean matrices. We see that the rank of the combined metrics where the latent parameters are different is roughly twice that of metric1, metric2, and their combination with identical row and column parameters across metrics. This is an important diagnostic, which can be used to ascertain if the model assumptions are (approximately) satisfied by the data in any multi-metric setting. In real-world case studies where the latent variables and mean matrices are unknown, the same analysis can be conducted using the singular value spectrums of the observed matrices. We point to Figure \ref{fig:singvals} for a specific instance of this diagnostic using real-world noisy observation data.

\begin{table}[h]
\centering
 \begin{tabular}{|| c | c ||} 
 \hline
Matrix & Approx. Rank\\ [0.5ex] 
 \hline\hline
metric1 & 9\\ 
 \hline
metric2 & 9 \\
 \hline
combined & 9 \\
(\textbf{same} row and column params) & \\
 \hline
combined & 17 \\
(\textbf{different} row and column params) & \\
 \hline
\end{tabular}
\caption{Ranks of the mean matrices for each metric and their concatenated matrices when the latent row and column parameters are held constant across metrics and otherwise.}
\label{table:diagnostic}
\end{table}

\section{Main Results} \label{sec:results}

Here, we present our main results, which bound the pre- and post-intervention prediction errors of our algorithm. We begin, however, with a crucial observation on the underlying mean tensor $\tM$, which justifies our algorithmic design. 

\begin{prop} \label{prop:linear_comb}
Assume Property \ref{property:low_rank} holds. Suppose the target unit is chosen uniformly at random amongst the $N$ units; equivalently, let
the units be re-indexed as per some permutation chosen uniformly at random. Then, with probability $1-r/N$, there exists a $\beta^* \in \Reals^{(N-1)}$ such that the target unit (represented by index $1$) satisfies
\begin{align} \label{eq:linear_comb}
	M_{1jk} &= \sum_{z=2}^{N} \beta^*_z \cdot M_{zjk}, 
\end{align}
for all $j \in [T]$ and $k \in [K]$. 
\end{prop}
Thus, under the low-rank property of $\tM$ (Property \ref{property:low_rank}), the target/treatment unit is shown to be a linear combination of the donor units across all metrics with high probability. This is the key property that is necessary in all synthetic control-like settings, and it allows us to flatten our third-order tensor into a matrix in order to utilize information across multiple metrics since the target unit is a linear combination of the donor pool across \textit{all metrics}.

More specifically, Proposition \ref{prop:linear_comb} establishes that for every metric $k \in [K]$, the first row of the lateral slice $\tM_{1, \cdot, k}$ is a linear combination of the other rows within that slice with high probability. Therefore, we can flatten $\tM$ into a $N \times kT$ matrix $\bM$ by concatenating the $K$ slices $\tM_{\cdot, \cdot, k}$ of $\tM$, and still maintain the linear relationship between the first row of $\bM$ and its other rows, i.e.,
\begin{align} \label{eq:flattened_mean_matrix}
	\bM &= [\tM_{\cdot, \cdot, 1}, \dots, \tM_{\cdot, \cdot, K}]. 
\end{align}
This linear relationship across all metrics allows us to combine the datasets from different metrics to effectively augment the pre-intervention period (Step 1 of Algorithm \ref{algorithm:mrsc}). For the rest of this exposition, let $\bM$ have the following SVD:
\begin{align}
	\bM &= \sum_{i=1}^r \tau_i \mu_i \nu_i^T,
\end{align}
where $\tau_i$ denote the singular values, and $\mu_i, \nu_i$ denote the left and right singular vectors, respectively. 

\subsection{Pre-intervention Error}
The following statement bounds the pre-intervention error of the mRSC algorithm; again, we remind the reader that $K=1$ reduces to the RSC framework.

\begin{theorem} \label{thm:pre_int}
Let the algorithmic hyper-parameter $\Delta = \bI$ (the identity matrix). Let $r$ and $\beta^*$ be defined as in \eqref{eq:low_rank_tensor} and \eqref{eq:linear_comb}, respectively. Suppose the following conditions hold:
\begin{enumerate}
	\item Properties \ref{property:low_rank}, \ref{property:boundedness}, \ref{property:noise.1} for some $\alpha \ge 1$, \ref{property:noise.2}, and \ref{property:masking_noise}. 
	
	\item The thresholding parameter $\lambda$ is chosen s.t. $\emph{rank}(\bhM) = r$. 
	
	\item $T_0 = \Theta(T)$. 
	
	\item The target unit, unit one, is chosen uniformly at random amongst the $N$ units. 
\end{enumerate}
Then with probability at least $1 - \frac{r}{N}$, 
\begin{align*} 
	\frac{1}{K}\sum_{k=1}^K \emph{MSE}_{T_0}(\hM_1^{(k)}) &\le \frac{4 \sigma^2 r}{KT_0} + C_1 C(\alpha) \, \| \beta^*\|_1^2 \, \frac{ \log^2(KNT_0)}{\rho^2 KT_0} \, \Bigg(r + \frac{\Big( (KT_0)^2 \rho + KNT_0 \Big) \log^3(KNT_0)}{\rho^2 \tau_r^2} \Bigg), 
\end{align*}
where $C_1 = (1 + \gamma + \Gamma + K_\alpha)^4$, $C(\alpha)$ is a positive constant that may depend on $\alpha \ge 1$, and $\tau_r$ is the $r$-th singular value of $\bM$. 
\end{theorem}

\subsection{Post-intervention Error}
We now proceed to bound the post-intervention MSE of our proposed algorithm. We begin, however, by stating the model class under consideration: 
\begin{align*}
	\mathcal{F} &= \{ \beta \in \Reals^{N-1}: \| \beta\|_2 ~\le B, \, \| \beta\|_0 ~\le r \},
\end{align*}
where $B$ is a positive constant. As is commonly assumed in generalization error analyses for regression problems, we consider candidate vectors $\beta \in \Reals^{N-1}$ that have bounded $\ell_2$-norm. Further, by Proposition 4 of \cite{asss}, if $\text{rank}(\bhM) = r$, then for any $\hbeta \in \Reals^{N-1}$, there exists a $\beta \in \Reals^{N-1}$ such that $\hbeta^T \bhM = \beta^T \bhM$ and $\| \beta \|_0 \le r$; hence, we can restrict our model class to $r$-sparse linear predictors. Combining the above observations, we consider the collection of candidate regression vectors within $\mathcal{F}$, i.e., the subset of vectors in $\Reals^{N-1}$ that have bounded $\ell_2$-norm and are $r$-sparse. 

\vspace{10pt}
\noindent \textit{Generating process.} Let us assume $\tM$ is generated as per a linear LVM. Specifically, for all $(i,j,k) \in [N] \times [T] \times [K]$,
\[ 
M_{ijk} = f(\theta_i, \rho_j, \omega_k) = \sum_{\ell = 1}^r \lambda_\ell \, \theta_{i, \ell} \, \rho_{j, \ell} \, \omega_{k, \ell},
\]
where $\lambda_\ell \in \Reals, \theta_i \in [0,1]^{d_1}, \rho_j \in [0,1]^{d_2}$, and $\omega_k \in [0,1]^{d_3}$ for some $d_1, d_2, d_3 \ge r$, and $r$ is defined as in \eqref{eq:low_rank_tensor}. 
Further, we make the natural assumption that $\theta_i, \rho_j$, and $\omega_k$ are sampled i.i.d. from some underlying (unknown) distributions $\Theta, P$, and $\Omega$, respectively. This gives rise to the following statement. 

\begin{theorem} \label{thm:post_int}
Let the conditions of Theorem \ref{thm:pre_int} hold. Further, let $\hbeta \in \mathcal{F}$.
Then for any metric $k \in [K]$, 
\begin{align} \label{eq:post_int_error}	
	\Ex \left[ \emph{MSE}_{T-T_0}(\hM_1^{(k)}) \right] &\le C_1 \Ex \left[ \frac{1}{K} \sum_{k'=1}^K \emph{MSE}_{T_0}(\hM_1^{(k')}) \right] + \frac{C_2 r^{3/2} \widehat{\alpha}^2}{\sqrt{KT_0}}   \| \beta^*\|_1.
\end{align}
Here, $C_1$ denotes an absolute constant, $C_2 = C B^2 \Gamma$ for some $C > 0$, and $\widehat{\alpha}^2 = \Ex[ \| \bhM \|_\max^2]$; lastly, we note that the expectation is taken with respect to the randomness in the data (i.e., measurement noise) and $\Theta, P, \Omega$.
\end{theorem}
In words, concatenating multiple relevant metrics effectively augments the pre-intervention period (number of training samples) by a factor of $K$. By Theorem \ref{thm:post_int}, this results in the post-intervention error decaying to zero by a factor of $\sqrt{K}$ faster, as evident by the generalization error term in \eqref{eq:post_int_error}. 
%
%
Further, we note that the first term on the righthand side of \eqref{eq:post_int_error}, $(1/K) \sum_{k'=1}^K \text{MSE}_{T_0}(\hM_1^{(k')})$, corresponds to the pre-intervention error across all $K$ metrics, as defined in Theorem \ref{thm:pre_int}. \\


\noindent{\bf Implications.} The statement of Theorem \ref{thm:pre_int} requires that the {\em correct} number of singular values are retained by the mRSC algorithm. In settings where all $r$ singular values of $\bM$ are roughly equal, i.e.,
\[ \tau_1 \approx \tau_2 \approx \dots \approx \tau_r = \Theta \Big( \sqrt{(KNT)/r} \Big),\]
 the pre-intervention prediction error vanishes as long as $KT_0$ scales faster than 
$\max(\sigma^2 r, \rho^{-4} r \log^5(N))$. Further, as long as $r = O(\log^{1/4} (N))$, the overall error vanishes with the same
scaling of $KT_0$. However, the question remains: how does one find a good $r$ in practice?

The purpose of the generalization error, such as that implied by Theorem \ref{thm:post_int}, is to precisely help resolve such a dilemma. Specifically, it 
suggests that the overall error is at most the pre-intervention (training) error plus a generalization error term that scales as $r^{3/2}/\sqrt{KT_0}$. Therefore, one should choose the
$r$ that minimizes this bound -- naturally, as $r$ increases, the pre-intervention error is likely to decrease, but the additional term $r^{3/2}/\sqrt{KT_0}$ 
will increase; therefore, a unique minima (in terms of the value of $r$) exists, and it can be found in a data driven manner.

\section{Experiments} \label{sec:experiments}

We establish the validity of our mRSC algorithm in three settings: 

\begin{enumerate}
\item \textbf{Idealized synthetic-data experiment}: using data generated by a known model (outlined in Section \ref{sec:model}), we conduct this experiment to empirically verify Theorems \ref{thm:pre_int} and \ref{thm:post_int}. In situations where the data-generating mechanism and {\em unobserved} means are known, we demonstrate that the mRSC algorithm outperforms the vanilla RSC algorithm \cite{rsc1} by achieving a lower forecasting prediction error. 

\item \textbf{Retail}: using Walmart sales data, we provide an instance of a ``real-world'' setting where mRSC outperforms RSC in forecasting future sales (the counterfactual). 

\item \textbf{Cricket}: considering the problem of forecasting scores in cricket, we first show that the trajectory of scores can be modeled as an instance of the mRSC model with multiple natural metrics of interest, and then, through extension experimentation, demonstrate the predictive prowess of the algorithm, which is also successful in capturing the nuances of the game.
\end{enumerate}

\subsection{Idealized synthetic-data experiment} \label{sec:mrsc-synthetic}

\paragraph{\bf Experimental setup.} We consider a setting where we have two metrics of interest, metricA and metricB. The data is generated as follows: we first sample sets of latent row and column parameters $\mathbb{S}_r, \mathbb{S}_c$, where $\mathbb{S}_r = \{s_k | s_k \sim \text{Uniform}(0, 1), 1 \leq k \leq 10 \}$ and  $\mathbb{S}_c = \{s_k | s_k \sim \text{Uniform}(0, 1), 1 \leq k \leq 10 \}$. We then fix the latent row and column parameters, $\theta_i$ and $\rho_j$, by sampling (with replacement) from $\mathbb{S}_r$ and $\mathbb{S}_c$. Note that $1 \leq i \leq N$ and $1 \leq j \leq T$, where $N$ and $T$ represent the dimensions of the matrices for each metric. In this experiment, we fix $T = 50$ and vary $N$ in the range $[50, 500]$.

For each metric, we use functions $f_a(\theta_i, \rho_j)$ and $f_b(\theta_i, \rho_j)$ to generate the mean matrix $\bM_a$ and $\bM_b$. Specifically, 
\begin{align*}
	f_a(\theta_i, \rho_j) &=  \frac{10}{1 + \exp(-\theta_i - \rho_j - (\alpha_a \theta_i \rho_j))},
\end{align*} 
where $\alpha_a = 0.7$. $f_b(\theta_i, \rho_j)$ is defined similarly but with $\alpha_b = 0.3$. We then let $\bM_a = [m_{a, ij}] = [f_a(\theta_i, \rho_j)]$ and $\bM_b = [m_{b, ij}] = [f_b(\theta_i, \rho_j)], \text{for } 1 \leq i \leq N, 1 \leq j \leq T$. 

Next, we generate the mean row of interest for each metric by using a fixed linear combination of the rows of the matrices $\bM_a$ and $\bM_b$, respectively. We append these rows of interest to the top of both matrices. We refer to them as rows $m_{a, 0}$ and $m_{b_0}$, respectively.

Independent Gaussian noise, $\mathbb{N}(0, 1)$ is then added to each entry of the matrices, $\bM_a$ and $\bM_b$, including the mean rows of interest at the top. This results in the observation matrices $\bY_a$ and $\bY_b$ for metricA and metricB. 

Given the $(N+1) \times T$ matrices $\bY_a$ and $\bY_b$, and an intervention time, $T_0 < T$, the goal is to estimate the unobserved rows $\hat{m}_{a, 0}$ and $\hat{m}_{b_0}$ in the post-intervention period, i.e., for columns $j$ where $j > T_0$. We achieve this by using estimates provided by the following:
\begin{enumerate}
\item \textbf{mRSC} algorithm presented in Section \ref{sec:algorithm}. This algorithm combines the two $(N+1) \times T$ matrices in to one matrix of dimensions $(N+1) \times 2T$. For the regression step we use equal weights for both metrics given the form of the generating functions $f_a$ and $f_b$.
\item \textbf{RSC} algorithm of \cite{rsc1} applied separately to $\bY_a$ and $\bY_b$ to generate estimates of each metric, independently.
\end{enumerate}
We conduct the experiment 100 times for each combination of $N$ and $T$ and average the resulting RMSE scores for the forecasts.

\begin{figure}
	\centering
	\includegraphics[width=0.4\textwidth]{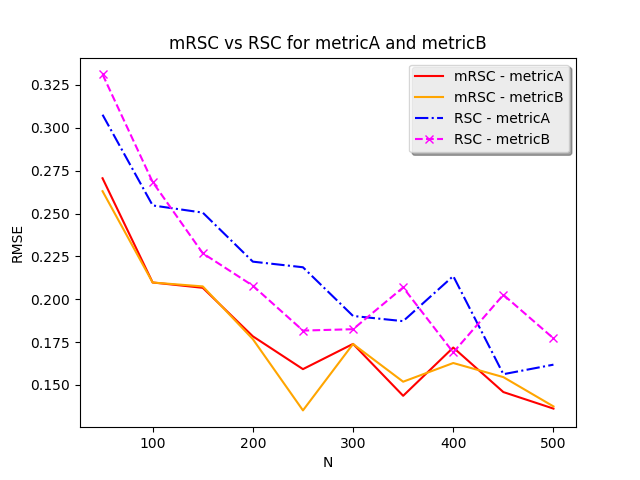}
	\caption{RMSE for mRSC and RSC algorithms for metricA and metricB for the experiment described in Section \ref{sec:mrsc-synthetic}.}
	\label{fig:mrsc-exp-synthetic}
\end{figure}

\paragraph{\bf Results.} Figure \ref{fig:mrsc-exp-synthetic} shows the results of the experiment. We note that for all levels of $N \in [50, 500]$, mRSC produces a lower RMSE value for the estimates for the first row of both metricA and metricB. This is perfectly in line with the expectations set by Theorems \ref{thm:pre_int} and \ref{thm:post_int}. Note that while the bounds in Theorem \ref{thm:post_int} improve by a factor of $\sqrt{2}$, it is an upper bound and we do not expect the RMSE values to necessarily shrink by that amount.

\subsection{Forecasting in retail} \label{sec:retail}
We consider the problem of forecasting weekly sales in retail. Here, we highlight a key utility of mRSC over RSC in the presence of sparse data. More specifically, our results demonstrate that when the pre-intervention period (training set) is short, then standard RSC methods fail to generalize well. On the other hand, by using auxiliary information from other metrics, mRSC effectively ``augments'' the training data, which allows it to overcome the difficulty of extrapolating from small sample sizes.

\paragraph{\bf Experimental setup.} We consider the Walmart dataset, which contains $T =143$ weekly sales information across $N = 45$ stores and $K = 81$ departments. We arbitrarily choose store one as the treatment unit, and introduce an ``artificial'' intervention at various points; this is done to study the effect of the pre-intervention period length on the predictive power for both mRSC and RSC methods. In particular, we consider the following pre-intervention points to be $15, 43$, and $108$ weeks, representing small to large pre-intervention periods (roughly $10\%, 30\%$, and $75\%$ of the entire time horizon $T$, respectively). Further, we consider three department subsets (representing three different metric subgroups): Departments $\{2, 5, 6, 7, 14, 23, 46, 55\}$, $\{17, 21, 22, 32, 55\}$, and $\{3, 16, 31, 56\}$. 


\paragraph{\bf Results.} In Table \ref{table:avg_mse}, we show the effect of the pre-intervention length on the RSC and mRSC's ability to forecast. In particular, we compute the average pre-intervention (training) and post-intervention (testing) MSEs across each of the three departmental subgroups (as described above) for both methods and for varying pre-intervention lengths. Although the RSC method consistently achieves a smaller average pre-intervention error, the mRSC consistently outperforms the RSC method in the post-intervention regime, especially when the pre-intervention stage is short. 
This is in line with our theoretical findings of the post-intervention error behavior, as stated in Theorem \ref{thm:post_int}; i.e., the benefit of incorporating multiple relevant metrics is exhibited by the mRSC algorithm's ability to generalize in the post-intervention regime, where the prediction error decays by a factor of $\sqrt{K}$ faster than that of the RSC algorithm.

\begin{figure}[H]
	\centering
	\subfigure[Dept. 56 (mRSC)]{\includegraphics[width=0.325\linewidth]{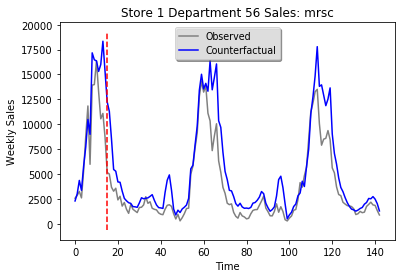}}
	\subfigure[Dept. 22 (mRSC)]{\includegraphics[width=0.325\linewidth]{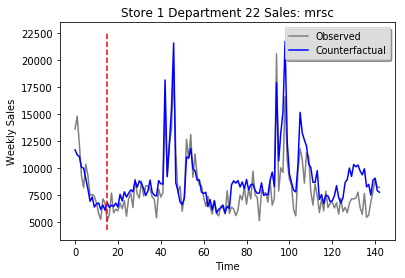}}	
	\\
	\subfigure[Dept. 56 (RSC)]{\includegraphics[width=0.325\linewidth]{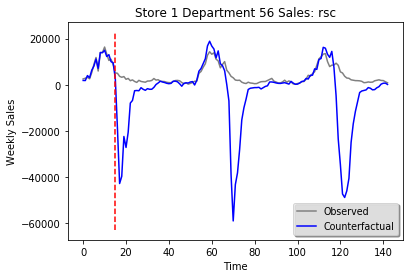}}
	\subfigure[Dept. 22 (RSC)]{\includegraphics[width=0.325\linewidth]{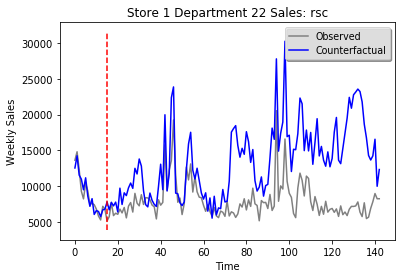}}	
	\caption{mRSC (top) and RSC (bottom) forecasts for departments 56 (left) and 22 (right) of store 1 using $T_0 = 15$ weeks.}
	\label{fig:walmart}
\end{figure}

We present Figure \ref{fig:walmart} to highlight two settings, departments 56 (left) and 22 (right), in which mRSC drastically outperforms RSC in extrapolating from a small training set ($T_0 = 15$ weeks). We highlight that the weekly sales axes between the subplots for each department, particularly department 56, are different; indeed, since the RSC algorithm was given such little training data, the RSC algorithm predicted negative sales values for department 56 and, hence, we have used different sales axes ranges to underscore the prediction quality gap between the two methods. 
As seen from these plots, the RSC
method struggles to extrapolate beyond the training period since the pre-intervention period is short. In general, the RSC method compensates for lack of data by overfitting to the pre-intervention observations and, thus,
misinterpreting noise for signal (as seen also by the smaller pre-intervention error in Table \ref{table:avg_mse}). Meanwhile, the mRSC overcomes this
challenge by incorporating sales information from other
departments. By effectively augmenting the pre-intervention period,
mRSC becomes robust to sparse data. However, it is worth noting that
both methods are able to extrapolate well in the presence of
sufficient data. 

\begin{table} 
\centering
\begin{tabular}{ c | c  c | c c }
\hline
 \multicolumn{1}{l}{}	     &	 \multicolumn{2}{c}{ Train Error ($10^6$)}  &	 \multicolumn{2}{c}{ Test Error  ($10^6$)} \\
\hline
 	$T_0$   &	 RSC 	&	mRSC  	&	 RSC	&	mRSC  	\\
 \hline
 10\%	&  {\bf 1.54} 	& 3.89		& 21.0		& {\bf 5.25} 	  \\
 30\%	&  {\bf 2.21}	& 3.51		& 19.4		& {\bf 4.62	}	  \\
 75\%	&  {\bf 4.22} 	& 5.33		& 3.32		& {\bf 2.48}  	\\
\hline 
 10\%	&  {\bf 0.67} 	& 2.61		& 14.4		& {\bf 2.48	}	  \\
 30\%	&  {\bf 0.79}	&1.21 		& 2.13		& {\bf 1.97	}	  \\
 75\%	&  {\bf 1.18} 	& 2.78		& 1.31		& {\bf 0.77	}	  \\
 \hline 
 10\%	&  {\bf 1.28} 	& 6.10		& 84.6		& {\bf 12.5	}	  \\
 30\%	&  {\bf 2.60}	& 3.45 		& {\bf 3.72	}	& 4.13	  \\
 75\%	&  {\bf 2.29} 	& 2.65		& 4.92		& {\bf 4.72}	  \\
\hline
\end{tabular}
\caption{Average pre-intervention (train) and post-intervention (test) MSE for RSC and mRSC methods.}
\label{table:avg_mse}
\end{table}

\subsection{Forecasting scores in cricket} \label{sec:cricket}
We now focus our attention on another real-world problem: forecasting scores in the game of cricket. A-priori, there is nothing obvious which indicates that this problem has any relationship to the model introduced in Section \ref{sec:model}. As an important contribution of this work, we show how cricket innings can be modeled as an instance of mRSC and we conduct extensive experimentation to demonstrate mRSC's excellent predictive performance.

As a starting point, we note that the mRSC setting is a natural candidate for modeling cricket because the game has two key metrics: {\em runs} and {\em wickets}. While the winner is the team that makes more total {\em runs}, the state of the game cannot be described by runs, alone. Wickets lost are as important as runs scored in helping determine the current state of the game. Additionally, it is well-understood among the followers and players of the game that the trajectory of runs scored {\em and} wickets lost are both crucial in determining how well a team can be expected to do for the remainder of the game. We formalize this intuition to model cricket innings and estimate the future total runs scored (and wickets lost) by a team using the mRSC algorithm.

In what follows, we first describe the most important aspects of the game of cricket. Next, we describe how to model an inning in cricket as an instance of mRSC. Finally, we use the mRSC algorithm to forecast scores in several hundred actual games to establish its predictive accuracy. In the process, we show that (a) both steps in the mRSC algorithm, i.e., de-noising and regression, are necessary when estimating the future scores in a cricket inning; (b) a learning algorithm that considers only runs, e.g., the RSC algorithm, would be insufficient and, thus, perform poorly because wickets lost are a critical component of the game; (c) somewhat counterintuitively, constraining the ``donor pool'' of innings to belong to the team of interest leads to poor predictive performance; and (d) using {\em real} examples, the mRSC model and algorithm are able to do justice in capturing the nuances of the game of cricket.

\subsubsection{A primer on cricket.}
Cricket is one of the most popular sports in the world. According to BBC, over a billion people tuned in on television 
to watch India play Pakistan in the ICC Cricket World Cup 2015 (\cite{bbccricket, wapocricket}). It is a fundamentally different 
game compared to some its most popular rivals. Unlike football (soccer), rugby, and basketball, cricket is an asymmetric game 
where both teams take turns to bat and bowl; and unlike baseball, cricket has far fewer innings, but each inning spans a 
significantly longer time. While cricket is played across three formats, the focus of this work is on the 50-over 
Limited Overs International (LOI) format, where each team only bats once and the winning team is the one that scores more 
total runs.

A batting inning in a LOI cricket game is defined by the total number of times 
the ball is bowled to a batter (similar to a ``pitch'' in baseball). In a 50-over LOI inning, the maximum number of balls bowled is $300$. 
A group of six balls is called an ``over''. Therefore, for the format under consideration, an inning can last up to $50$ overs or $300$ balls. 
Each batting team gets a budget of $10$ ``outs'', which are defined as the number of times the team loses a batter. In cricket, a batter 
getting ``out'' is also referred to as the team having lost a ``wicket''. Therefore, the state of a cricket inning is defined by the tuple of 
({\em run scores}, {\em wickets lost}) at each point in the inning. We shall call the trajectory of an inning by a sequence of such tuples. In the event that the team loses all $10$ of its wickets before the maximum number of balls (300) are bowled, 
the innings ends immediately. The team batting second has to score one more run than the team batting first to win the game. 
As an example, in the ICC Cricket World Cup Final in 2015, New Zealand batted first and lost all ten wickets for a cumulative score of $183$ in 
45 overs (270 balls). Australia were set a target of 184 in a maximum of 50 overs but they chased it successfully in 33 overs and 1 ball (199 balls) 
for the loss of only 3 wickets.

\subsubsection{The forecasting problem.}
We consider the problem of {\em score trajectory forecasting}, where the goal
is to estimate or predict the (score, wicket) tuple for all the {\em future} remaining balls of an ongoing inning. We wish to develop an entirely data-driven approach to address this problem while avoiding the complex models employed in some of the prior works (see Section \ref{sec:related}). Just as in the rest of this work, we are using the mRSC setting as a vehicle 
for {\em predictions}.

\subsubsection{An mRSC model.}

Given that the problem involves twin-metrics of runs and wickets, the vanilla RSC algorithm would be unsatisfactory in this setting. We cast this problem as an instance of the mRSC model outlined in Section \ref{sec:model}. However, it is not obvious how an inning in cricket can be reduced to an instance of the mRSC problem. We proceed to show that next.

\smallskip
\noindent{\bf Intuition.} 
It is important to establish the validity of the {\em latent variable model} and, hence, the factor model or low-rank tensor 
model in this setting. It is motivated by the following minimal property that sports fans and pundits expect to hold 
for a variety of sports: the performance of a team is determined by (1) the {\em game context} that remains constant through 
the duration of the game and (2) the {\em within innings context} that changes during the innings. In cricket, the
{\em game context} may correspond to the players, coach, the mental state of the players, 
the importance of the game (e.g., world cup final), the game venue and fan base, the pitch or ground 
condition, etc. At the same time, the {\em within innings context} may correspond to stage of the innings, power play, etc. Next, we 
formalize this seemingly universal intuition across sports in the context of cricket. 

\smallskip
\noindent{\bf Formalism.}  The performance, as mentioned before, is measured by the runs scored and the 
wickets lost. To that end, we shall index an inning by $i$, and ball within the inning by $j$. Let there be a total 
of $n$ innings and $b$ balls (within LOI, $b = 300$). Let $X_{ij}$ and $W_{ij}$ denote the runs scored and 
wickets lost on ball $j$ within inning $i$. We shall model $X_{ij}$ and $W_{ij}$ as independent random 
variables with mean $m_{ij} = {\mathbb E}[X_{ij}]$ and $\lambda_{ij} = {\mathbb E}[W_{ij}]$. Further, 
let 
\begin{align}
m_{ij} & = f_{r}(\theta_i, \rho_j) \equiv f(\theta_i, \rho_j, \omega_r), \\
\lambda_{ij} & = f_{w}(\theta_i, \rho_j) \equiv  f(\theta_i, \rho_j, \omega_w)
\end{align}
for all $i \in [n], j \in [b]$\footnote{We use notation $[x] = \{1,\dots, x\}$ for integer $x$.} where 
{\em latent} parameter $\theta_i \in \Omega_1 = [r_1, s_1]^{d_1}$ captures the {\em game context};
{\em latent} parameter $\rho_j \in \Omega_2 = [r_2, s_2]^{d_2}$ captures the {\em within innings context}; 
{\em latent} functions $f_{r}, f_w: \Omega_1 \times \Omega_2 \to [0,\infty)$ capture 
complexity of the underlying model for runs scored and wickets lost; and these functions
are thought be coming from a class of parametric functions formalized 
as $f_r(\cdot, \cdot) \equiv f(\cdot, \cdot, \omega_r)$ and 
$f_w(\cdot, \cdot) \equiv f(\cdot, \cdot, \omega_w)$ for some parameters $\omega_r, \omega_w \in \Omega_3 = [r_3, s_3]^{d_3}$.
For simplicity and without loss of generality, we shall assume that $r_1=r_2 = r_3 = 0$,
$s_1 = s_2 =s_3= 1$, $d_1 = d_2 = d_3 = d \geq 1$. Now this naturally fits the Latent variable model
or factor model we have considered in this work. Therefore, in principle, we can apply mRSC approach
for forecasting run / wicket trajectory.

\smallskip
\noindent{\bf Diagnostic.} 
Proposition \ref{prop:low_rank_approx} suggests that if indeed our model assumption holds in practice, then the
data matrix of score and wicket trajectories and their combination ought to be well approximated by a low-rank matrix. Moreover, for the mRSC model to hold, we expect the ranks of all these matrices to be approximately equal, as discussed in Section \ref{sec:falsifiability}. For the low-rank to manifest, each matrix's spectrum should be concentrated on top few principal components. This gives us way to test falsifiability of our model for the game of cricket, and more generally any game of interest.

\begin{figure}
	\centering
	\includegraphics[width=0.4\textwidth]{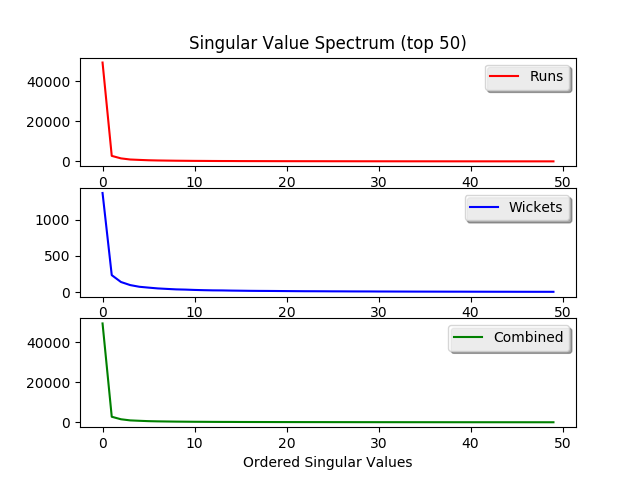}
	\caption{The Singular Value spectrum for all innings in the dataset (runs, wickets and combined matrices of dimensions 4700 $\times$ 300. Only showing the top 50 singular values, in descending order.}
	\label{fig:singvals}
\end{figure}

In the context of cricket, we consider a matrix comprising 4700 LOI innings from 1999 to 2017, i.e. as many rows and $300$
columns. Figure \ref{fig:singvals} shows the spectrum of the top 50 singular values (sorted in descending order) for each matrix. Note that the magnitude of the singular values is expected to different between the two metrics owing to the assumed differences between the generating functions, $f_r$ and $f_w$. It is the relative proportions between the various singular values that we are interested in. The plots clearly support the implications that most of the spectrum is concentrated within the top few ($8$) principal components. The shapes for all three spectrums are also nearly identical, the magnitude differences notwithstanding. Indeed, we note that the over 99.5\% of the ``energy'' in each matrix is captured by the top 8 singular values. We determine this by calculating the ratio of the sum of squares of the top 8 singular values to the sum of squares of all singular values of each matrix.

\subsubsection{Experiments and Evaluation} \label{sec:experimentalsetup}
We describe the details of the experiments to evaluate the performance of the algorithm.

\smallskip
\noindent{\bf Data.} For all experiments described in the rest of this section, we consider 750 most recent LOI innings spanning seven years (2010-2017) and introduce an arbitrary intervention at the 30 over mark (180 balls), unless stated otherwise. For each inning, forecasts for the entire remainder of the innings (i.e., 20 overs (or 120 balls)) are estimated using our mRSC algorithm. For the donor pool, we take into consideration all LOI (first) innings from the year 1999 onwards, which comprises about 2400 innings.

\smallskip
\noindent{\bf Objective.} Even though the twin metrics, runs scored and wickets lost, are equally important in helping to determine the current state of the game and in forecasting the future, the key evaluation metric we will consider is the runs scored. We focus on this particular metric since the winner of game is solely a function of the number of runs scored. i.e., if team A scores more runs than team B, then team A is declared the winner if even team A has lost more wickets. Therefore, while we use {\em both} runs and wickets to train the mRSC model, all evaluations will only focus on the total number of runs scored. 

\smallskip
\noindent {\bf Evaluation metrics.} Given that the ground truth (i.e., the mean trajectory of runs scored and wickets lost) are latent and unobserved, we use the {\em actual} observations from each of these innings to measure our forecasting performance. We will use two metrics to statistically evaluate the performance of our forecast algorithm:

\smallskip
\noindent {\em MAPE.} We split the run forecast trajectories into four durations of increasing length: 5 overs (30-35 overs), 10 overs (30-40 overs), 15 overs (30-45 overs), and 20 overs (30-50 overs). Next, we compute the mean absolute percentage error (MAPE) in the range $[0, 1]$ for each inning and forecast period. MAPE helps us quantify the quality of runs-forecasts. A distribution of these errors, along with the estimated mean and median values, are reported.

\smallskip
\noindent {\em $R^2$.} To statistically quantify how much of the variation in the data is captured by our forecast algorithm, we compute the $R^2$ statistic for the forecasts made at the following overs: $\{35, 40, 45\}$. For each of these points in the forecast trajectory, we compute the $R^2$ statistic over all innings considered. The baseline for $R^2$ computations is the sample average (runs) of all innings in the donor pool at the corresponding overs, i.e., $\{35, 40, 45\}$. This is akin to the baseline used in computing the $R^2$ statistic in regression problems, which is simply the sample average of the output variables in the training data. 

\smallskip
\noindent {\bf Comparison.} In order to establish the efficacy of the mRSC algorithm, we conduct the following comparisons to study the importance of each facet of the mRSC algorithm: 
\begin{enumerate}
\item {\bf Criteria 1: value of the WLS}. comparison with an algorithm that uses the de-noised donor pool matrix (result of step 1 of the algorithm), but takes the sample average of the donor pool runs instead of a weighted least squared regression;

\item {\bf Criteria 2: value of de-noising}. comparison with an algorithm that ignores the de-noising step (step 1 of the algorithm) and performs a WLS regression directly on the observed data in the donor pool;

\item {\bf Criteria 3: value of using all data (Stein's Paradox)}. comparison with an algorithm that restricts the donor pool to belong to past innings played by the {\em same} team for which we are making the prediction. It is commonly assumed that in order to predict the future of a cricket innings, the ``context'' is represented by the identity of the team under consideration. This was also assumed about baseball and refuted (ref. Stein's Paradox \cite{stein1}). Filtering the donor pool to belong to innings played by the same team will allow us to test such an assumption.

\item {\bf Criteria 4: value of wickets}. we look for examples of actual innings in cricket where considering a single metric (e.g., runs) would not properly do justice to the state of the innings and one would expect runs-based forecasts from the vanilla RSC algorithm to be inferior to the mRSC algorithm, which is capable of taking the loss of wickets into account as well.

\item {\bf Criteria 5: value of our model}. finally, and perhaps most significantly, we compare the algorithm's forecasts with the context and details within some selected actual games/innings. Such a comparison against actual game context and cricketing intuition will help determine whether our algorithm (and by extension, the mRSC model) was able to capture the nuances of the game. 
\end{enumerate}

\subsubsection{Statistical Performance Evaluation}
\begin{figure}
	\centering
	\includegraphics[width=0.4\textwidth]{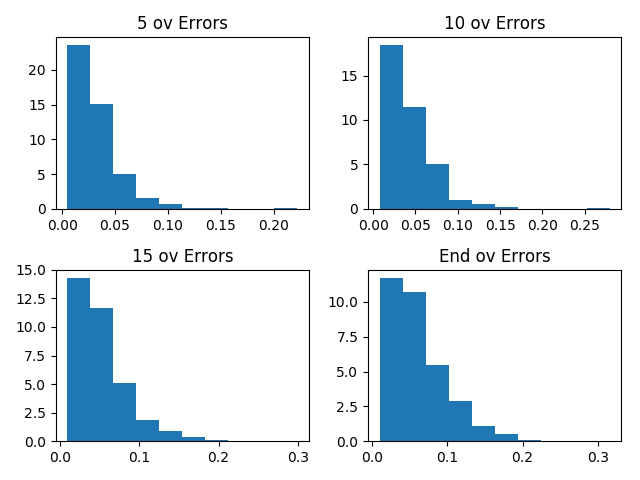}
	\caption{MAPE forecast error distributions for interventions made at the 30 over (180 balls) mark. Algorithm uses the top 5 singular values.}
	\label{fig:mapehist1}
\end{figure}

Figure \ref{fig:mapehist1} shows the distribution of MAPE statistic for the runs-forecasts from our algorithm for each of the four forecasts: 5 over, 10 over, 15 over, and 20 over periods with the intervention at the 30th over mark. The clear left-skew hints at most errors being small. The MAPE statistic is larger the longer the forecast horizon considered, which is to be expected. The median MAPE from our algorithm over the longest forecast horizon (20 overs) is only about 5\% while the shortest-horizon forecasts have a median of about 2.5\%.  In the cricket context, this should be considered excellent. Table \ref{table:mape1} notes the mean and median for the distribution of the MAPE statistic for our algorithm in comparison to the de-noised donor-pool averaging algorithm described in Criteria 1 in Section \ref{sec:experimentalsetup}. Our algorithm outperforms the donor-pool averaging algorithm for all forecast intervals on both the mean and median MAPE statistics. This establishes the clear value of Step 2 (regression) of the mRSC algorithm.
\begin{table}[h!]
\centering
 \begin{tabular}{|| c | c c | c c ||} 
 \hline
 Forecast	 &  \multicolumn{2}{c |}{mRSC Algorithm}  &  \multicolumn{2}{c ||}{Donor-Pool Avg}  \\
Interval & Mean & Median & Mean & Median \\ [0.5ex] 
 \hline\hline
 5 overs & 0.033 & 0.027 & 0.169 & 0.124\\ 
 \hline
 10 overs & 0.043 & 0.037 & 0.169 &  0.126\\
 \hline
 15 overs & 0.053 & 0.043 & 0.170 & 0.124 \\
 \hline
 20 overs & 0.062 & 0.051  & 0.173 & 0.121\\
 \hline
\end{tabular}
\caption{Forecasts from the mRSC algorithm vs the donor pool averaging algorithm. MAPE forecast error means and medians for interventions made at the 30 over (180 balls) mark. Algorithm uses the top 7 singular values.}
\label{table:mape1}
\end{table}

Tables \ref{table:r2-1} and \ref{table:r2-2} report the $R^2$ statistics for the 5, 10, and 15 over forecasts across all innings produced by our algorithm in comparison to an algorithm that performs the WLS regression step without de-noising (see Criteria 2 in Section  \ref{sec:experimentalsetup}). Note that the intervention is at the 10th over mark in Table \ref{table:r2-1} and at the 30th over mark in Table \ref{table:r2-2}. Both tables show that the de-noising step can only be an advantage, as originally argued in \cite{rsc1}. However, the advantage of the de-noising step is more pronounced in situations with less training data, i.e., an earlier intervention. This is to be expected and is perfectly in line with the other experiments conducted in this work. The $R^2$ values indicate that our algorithm is able to capture most of the variability in the data and that there is significant benefit in using the mRSC framework. The decline in median $R^2$ values as we increase the forecast horizon from 5 overs (30 balls) to 15 overs (90 balls) is to be expected because the forecast accuracy degrades with the length of the forecast horizon. 

\begin{table}[h!]
\centering
 \begin{tabular}{|| c | c c c ||} 
 \hline
 Intervention: 30th ov & 35 ov & 40 ov & 45 ov\\ [0.5ex] 
 \hline\hline
 mRSC Algorithm & 0.924 & 0.843 & 0.791\\ 
 \hline
Regression (noisy) & 0.921 & 0.839 & 0.787 \\
 \hline
\end{tabular}
\caption{$R^2$ of the forecasts from the mRSC algorithm vs the one where no donor-pool de-noising is performed as a precursor to the WLS regression step. Forecasts at the 35th, 40th, and 45th overs with the intervention at the 30 over (180 ball) mark. Algorithm uses the top 7 singular values.}
\label{table:r2-1}
\end{table}

\begin{table}[h!]
\centering
 \begin{tabular}{|| c | c c c ||} 
 \hline
 Intervention: 10th ov & 15 ov & 20 ov & 25 ov\\ [0.5ex] 
 \hline\hline
 mRSC Algorithm & 0.69 & 0.39 & 0.15\\ 
 \hline
 Regression (noisy) & 0.67 & 0.34 & 0.07\\
 \hline
\end{tabular}
\caption{$R^2$ of the forecasts from the mRSC algorithm vs the one where no donor-pool de-noising is performed as a precursor to the WLS regression step. Forecasts at the 15th, 20th, and 25th overs with the intervention at the 10 over (60 ball) mark. Algorithm uses the top 7 singular values.}
\label{table:r2-2}
\end{table}

\subsubsection{Stein's Paradox in cricket} \label{sec:donorpooleval}
We now show that the famous statistical paradox attributed to Charles Stein also makes its appearance in cricket. Using baseball as an example, Stein's paradox claims that the better forecasts of the future performance of a batter can be determined by considering not just the batter's own past performance but also that of other batters even if their performance might be independent of the batter under consideration \cite{stein1}. We use Table \ref{table:keyinsight2} to summarize a similar fact in cricket, using our forecast algorithm. Instead of considering all past innings in the donor pool, we use a donor pool comprising innings {\em only} from the same team as the one that played the innings we are forecasting. Table \ref{table:keyinsight2} shows that the median of MAPE statistics over all forecast horizons are $\sim$ 3x lower when using all past innings in the donor pool compared to a team-specific donor pool. This experiment also validates our latent variable model which allows a complex set of latent parameters to model cricket innings.

\begin{table}[h!]
\centering
 \begin{tabular}{|| c | c |  c c c c ||} 
 \hline
Team	 &  Donor Pool & 5ov & 10 ov   & 15 ov &  20 ov \\[0.5ex] 
 \hline\hline
England & All innings & \textbf{0.025} & \textbf{0.031} & \textbf{0.039} & \textbf{0.043} \\ 
 & Restricted & 0.065 & 0.097 & 0.113 & 0.127\\ 
 \hline
 Pakistan & All innings & \textbf{0.029} & \textbf{0.039} & \textbf{0.046} & \textbf{0.052} \\ 
 & Restricted & 0.072 & 0.104 & 0.113 & 0.119 \\ 
 \hline
  India & All innings & \textbf{0.027} & \textbf{0.037} & \textbf{0.045} & \textbf{0.049} \\ 
 & Restricted & 0.067 & 0.093 & 0.101 & 0.108 \\ 
  \hline
  Australia & All innings & \textbf{0.027} & \textbf{0.040} & \textbf{0.044} & \textbf{0.053} \\
 & Restricted & 0.063 & 0.084 & 0.094 & 0.104 \\ 
 \hline
\end{tabular}
\caption{MAPE medians for innings played by specific teams on a donor pool with all innings compared to one which is restricted to only include past innings from the same team. The rest of the algorithm is exactly the same for both. The intervention takes place at the 30 over mark and the forecast horizons are 5, 10, 15, and 20 overs.}
\label{table:keyinsight2}
\end{table}

\subsubsection{Capturing the effect of wickets: comparison to RSC} \label{sec:comparisonRSC}
We first consider the Champions Trophy 2013 game between Sri Lanka and Australia, played on June 17, 2013 at the Kennington Oval in London. In the second innings, Australia were chasing 254 for victory in their maximum allocation of 50 overs. However, given the tournament context, Australia would have to successfully chase the target in 29 overs to have a shot at qualifying for the semi-finals. In their attempt to reach the target as quickly as possible, Australia played aggressively and lost wickets at a higher rate than normal. They couldn't qualify for the semi-finals but also fell short of the target altogether, and  were bowled out for 233. At the 30th over mark, Australia's scorecard read 192 runs scored. Purely based on runs scored, that is a high score and an algorithm which takes {\em only} runs in to account, e.g. the vanilla RSC algorithm, would likely forecast that Australia would overhaul their target of 254 very comfortably. Indeed, the RSC algorithm forecasts that Australia would have reached their target in the 39th over. However, by that 30th over mark Australia had lost 8 wickets. This additional context provided by wickets lost is not captured by the RSC algorithm but is crucial in understanding the state of the game. In cricketing terms 192-8 in 30 overs would not be considered an advantageous position when chasing a target of 254. It is more likely for a team to get bowled out, i.e., lost all 10 of their wickets, before reaching their target. We use the mRSC algorithm to study the effect of effects. Using cross-validation, we find that the best ratio of objective function weights between runs and wickets is 1:4. Using this ratio of weights, the mRSC algorithm forecasts that Australia would be all out, i.e., lose all 10 wickets, and make 225 runs. This is very close to the actual final score of 233 all out. In cricketing terms, that is a far better forecast than that provided by the RSC algorithm which made little cricketing sense. 

We now consider the game played between England and New Zealand in Southhamption on June 14, 2015. Batting first, England were looking good to reach a score in excess of 350 by the 41th over when they had only lost 5 wickets but already scored 283 runs. However, from being a position of ascendency, their fortunes suddenly dipped and they ended up losing all five remaining wickets within the next four overs to get bowled out for 302. While it is hard to imagine any algorithm forecasting the sudden collapse, we use the 43rd over as a benchmark to compare the two algorithms. At that stage, England had progressed to 291 runs but lost 7 wickets. In cricketing terms, that is no longer a situation where one of would expect the team to cross 350 like it appeared only two overs back. The RSC algorithm does not grasp the change in the state of the game purely based on runs scored. It projects that England would still go on to make 359 runs. The mRSC algorithm, using a runs: wickets weights ratio of 1:2, forecasts a final score of 315 all-out. Once again, we notice that the mRSC algorithm is able to capture the effect of wickets, which is crucial in the game of cricket, to produce more accurate and {\em believable} forecasts.

\subsubsection{Capturing cricketing sense: case studies} \label{sec:forecastcasestudies}
\begin{figure}[H]
	\centering
	\subfigure[25 over marks]{\label{fig:ind-aus-1-25}\includegraphics[width=0.325\textwidth]{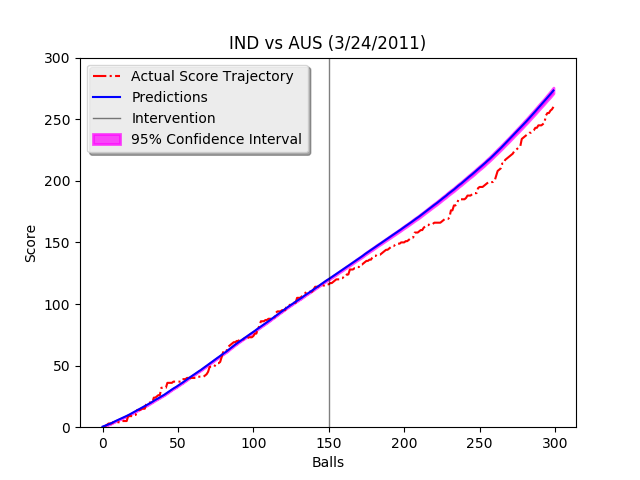}}
	\subfigure[40 over marks]{\label{fig:ind-aus-1-40}\includegraphics[width=0.325\textwidth]{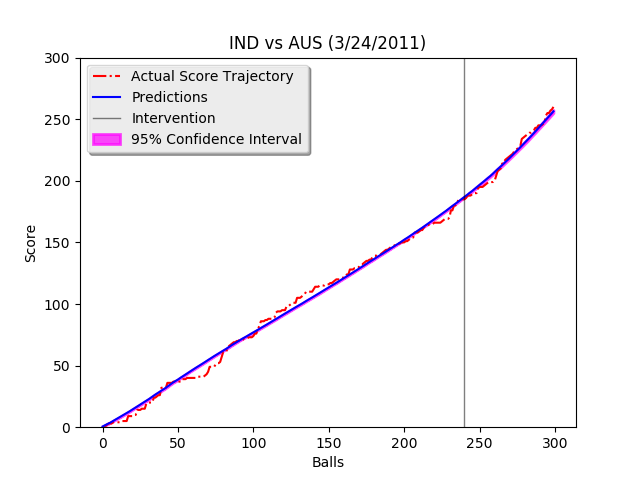}}
	\subfigure[45 over marks]{\label{fig:ind-aus-1-45}\includegraphics[width=0.325\textwidth]{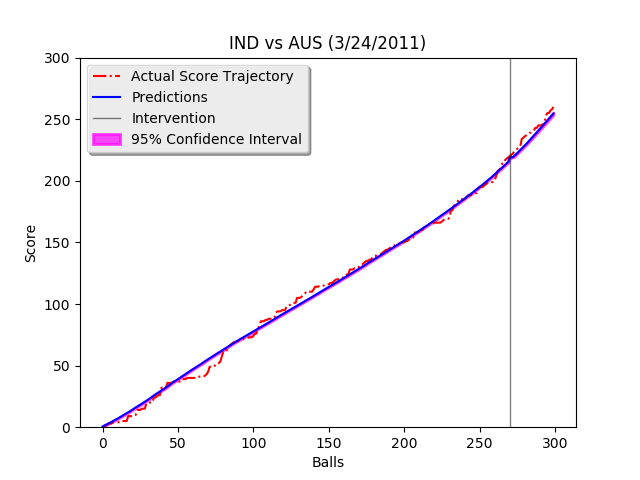}}
	\caption{India vs Aus (WC 2011). First Innings (Australia batting). Interventions at the 25, 40 and 45 over marks. Actual and Forecast trajectories with the 95\% uncertainty interval.}
	\label{fig:aus-ind-1}
\end{figure}

We use examples from actual games to highlight important features of our forecast algorithm. \\ \\
\noindent
\textbf{India vs Australia, World Cup Quarter Final 2011}. We use the ICC World Cup 2011 Quarter Final game between India and Australia. Batting first, Australia made 260 runs for the loss of 6 wickets in their allocated 50 overs. At the 25 over mark, Australia had made 116 runs for the loss of only two wickets. Australia's captain and best batsman, Ricky Ponting, was batting and looked set to lead Australia to a final score in excess of 275-280. However, India were able to claw their way back by taking two quick wickets between overs 30-35 and slowed Australia's progress. Eventually, that slowdown cost Australia a vital few runs and they ended with a good final score of 260--but it could have been better. Figure \ref{fig:aus-ind-1} shows the actual innings and trajectory forecasts produced by our algorithm for interventions at the 25, 40 and 45 over marks. Notice that Figure \ref{fig:ind-aus-1-25} shows that our algorithm forecasted Australia would make more runs than they eventually ended up with. This is what one would expect to happen--we forecast based on the current state of the innings and not using future information. Note that the forecast trajectories at the 40 and 45 over interventions match exceptionally well with reality now that the algorithm has observed India's fightback. This is an important feature of the algorithm and it happens because the pre-intervention fit changes for each of the three points of intervention. The events from overs 30-35 play a significant role in estimating a {\em different} synthetic control, $\widehat{\beta}$, for the interventions at over 40 and 45 compared to the intervention at over 25 (when those events were yet to happen). 

Another feature worthy of highlighting is that the algorithm is able to project the rise in the rate of scoring towards the end of an innings--this is a well-known feature of cricket innings and any data-driven algorithm should naturally be able to bring this to light. Finally, note the smoothness of the forecasts (in blue) compared to observations (in red). This is a key feature of our algorithm, which ``de-noises'' the data matrix to retain only the top few singular values to estimate the mean future trajectory. 



\begin{figure}[H]
	\centering
	\subfigure[35 over marks]{\label{fig:ind-aus-2-25}\includegraphics[width=0.325\textwidth]{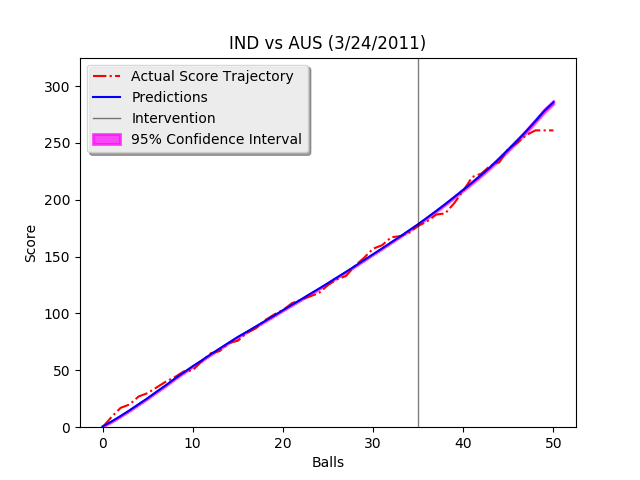}}
	\subfigure[40 over marks]{\label{fig:ind-aus-2-40}\includegraphics[width=0.325\textwidth]{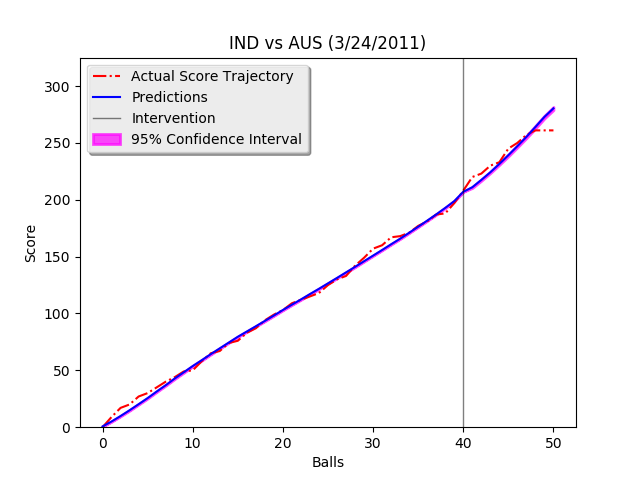}}
	\subfigure[45 over marks]{\label{fig:ind-aus-2-45}\includegraphics[width=0.325\textwidth]{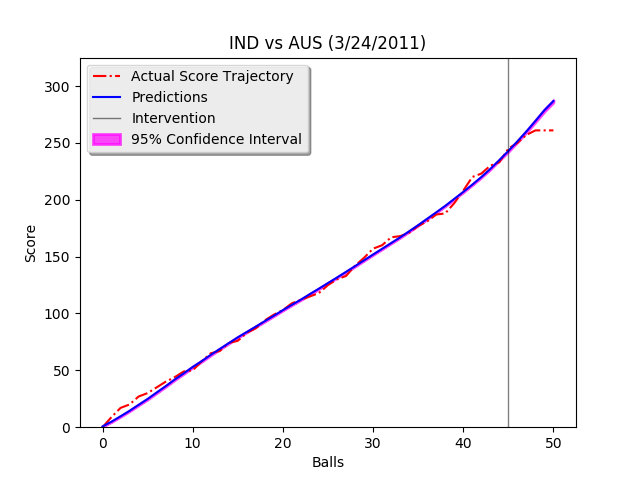}}	
	\caption{India vs Aus (WC 2011). Second Innings (India batting). Interventions at the 35, 40 and 45 over marks. Actual and Forecast trajectories with the 95\% uncertainty interval.}
	\label{fig:aus-ind-2}
\end{figure}

We now look at the second inning. Here, India was able to chase the target down and win the game with relative ease. 
Figure \ref{fig:aus-ind-2} shows the forecasts at the intervention points of 35, 40, and 45 overs for India's innings. The forecast trajectories are exceptionally close to reality and showcase the predictive ability of the algorithm. Once again notice the rise in score rate towards the late stages of the innings, similar to the first innings. We note that the flatlining of the actual score in the innings (red line) is simply due to the fact that India had won the game in the 48th over and no more runs were added. \\

\noindent
\textbf{Zimbabwe vs Australia, Feb 4 2001.} Zimbabwe and Australia played a LOI game in Perth in 2001. Australia, world champions then, were considered too strong for Zimbabwe and batting first made a well-above par 302 runs for the loss of only four wickets. The target was considered out of Zimbabwe's reach. Zimbabwe started poorly and had made only 91 for the loss of their top three batsmen by the 19th over. However, Stuart Carlisle and Grant Flower combined for a remarkable partnership to take Zimbabwe very close to the finish line. Eventually, Australia got both batsmen out just in the nick of time and ended up winning the game by just one run. We show the forecast trajectories at the 35, 40 and 45 over marks--all during the Carlisli-Flower partnership. The forecasts track reality quite well. A key feature to highlight here is the smoothness of the forecasts (in blue) compared to reality (in red). This is a key feature of our algorithm which ``de-noises'' the data matrix to retain only the top few singular values. The resulting smoothness is the mean effect we are trying to estimate and it is no surprise that the forecast trajectories bring this feature to light.

\begin{figure}[H]
	\centering
	\subfigure[35 over marks]{\label{fig:zim-aus-2-35}\includegraphics[width=0.325\textwidth]{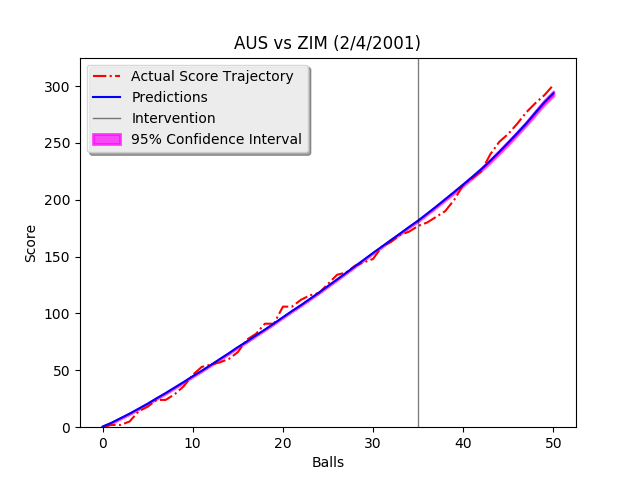}}
	\subfigure[40 over marks]{\label{fig:zim-aus-2-40}\includegraphics[width=0.325\textwidth]{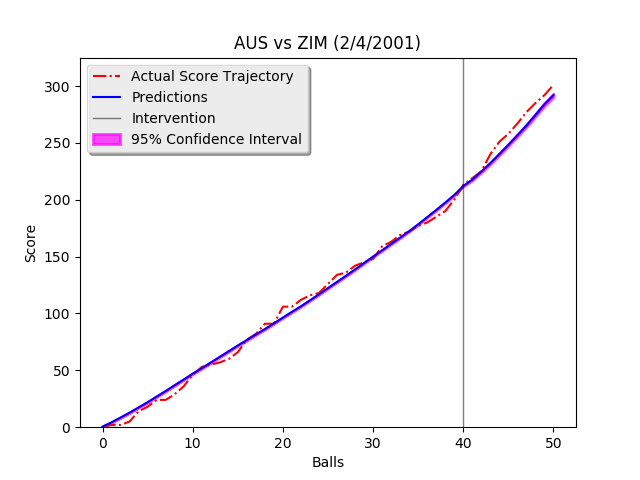}}
	\subfigure[45 over marks]{\label{fig:zim-aus-2-45}\includegraphics[width=0.325\textwidth]{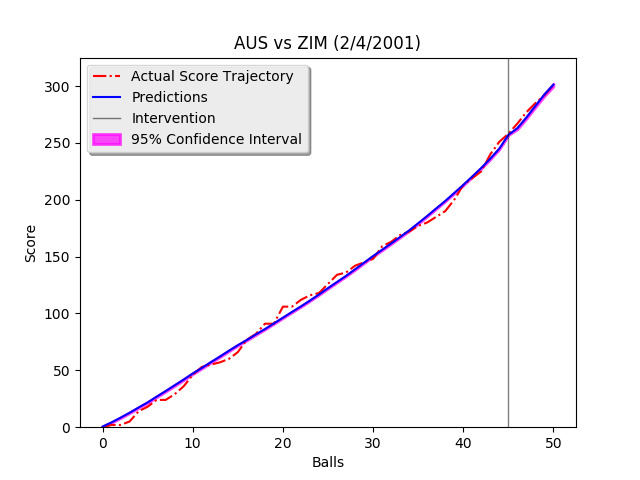}}
	\caption{Zimbabwe vs Aus (2001). Second Innings (Zimbabwe batting). Interventions at the 35 over, 40 over and 45 over mark.}
	\label{fig:aus-zim-2}
\end{figure}

\section{Conclusion}


{\bf Summary.} In this work, we focus on the problem of estimating the (robust) synthetic control and using it to forecast the future metric measurement evolution for a unit of interest under the assumption that a potential intervention has no statistical effect. Synthetic control (SC) \cite{abadie1, abadie2, abadie3}, robust synthetic control (RSC) \cite{rsc1}, and its variants perform poorly when the training (pre-intervention) data is too little or too sparse. We introduce the multi-dimensional robust synthetic control (mRSC) algorithm, which overcomes this limitation and generalizes the RSC algorithm. This generalization allows us to present a natural and principled way to include multiple (related) metrics to assist with better inference. The latent variable model lies at the heart of the mRSC model and is a natural extension of the factor model, which is commonly assumed in the SC literature. Our algorithm exploits the proposed low-rank tensor structure to ``de-noise'' and then estimate the (linear) synthetic control via weighted least squares. 
This produces a consistent estimator where the post-intervention (testing) MSE decays to zero by a factor of $\sqrt{K}$ faster than the RSC algorithm. 

Through extensive experimentation using synthetically generated datasets and real-world data (in the context of retail), we confirm the theoretical properties of the mRSC algorithm. Finally, we consider the problem of forecasting scores in the game of cricket to illustrate the modeling prowess and predictive precision of the mRSC algorithm.

\vspace{10pt}
\noindent {\bf Forecasting vis-\'{a}-vis Synthetic Control.}
While observational studies and randomized control trials are all concerned with estimating the unobserved counterfactuals for a unit of interest that experiences an intervention, how does one determine the performance of a counterfactual estimation method without access to the ground-truth? Although it is possible to use the pre-intervention data to cross-validate the performance of any estimation method, such a methodology ignores the period of interest: the post-intervention period. An alternate and more effective approach is to study the performance of an estimation method on units that do not experience the intervention, i.e., the {\em placebo} units. If the method is able to accurately estimate the observed post-intervention evolution of the placebo unit(s), it would be reasonable to assume that it would perform well in estimating the unobserved counterfactuals for the unit of interest. Therefore, in this work, we focus on evaluating the estimates of many such placebo units to establish the efficacy of our proposed method. Additionally, given that there is a temporal ordering of the data with clearly defined pre- and post- intervention period(s), our post-intervention estimation problem is akin to a forecasting problem; for units which do not experience any intervention, our goal is to accurately estimate the future and our estimates are evaluated against the observed data. This post-intervention period forecast accuracy becomes our primary metric of comparison and evaluation.

Given the discussion above, the method presented in this work serves a dual purpose: (a) it can be used as a method to estimate the (synthetic) control for a unit of interest that experiences an intervention; (b) as long as the temporal or sequential dimension is relative and not absolute, it can be used as a method to forecast the future evolution of any unit of interest. More precisely, this is only possible when the donor units have observations for both the past (pre-intervention period) {\em and} the future (post-intervention period), e.g., in the game of cricket, the donor pool comprises of a large set of already completed innings. Therefore, the mRSC method presented in this work is not a general time series forecasting algorithm: it requires the future to be known for the donor units, which can then assist in estimating the counterfactual for the unit of interest. For more details on the contrast between related work on time series forecasting and the synthetic control based algorithm presented in this work, we refer the reader to Section \ref{sec:related}.

\bibliographystyle{ACM-Reference-Format}
\bibliography{mainbib.bib} 

\begin{thebibliography}{26}
\providecommand{\natexlab}[1]{#1}
\providecommand{\url}[1]{\texttt{#1}}
\expandafter\ifx\csname urlstyle\endcsname\relax
  \providecommand{\doi}[1]{doi: #1}\else
  \providecommand{\doi}{doi: \begingroup \urlstyle{rm}\Url}\fi

\bibitem[kag(2014)]{kaggle-walmart}
Kaggle: Walmart recruiting - store sales forecasting, 2014.
\newblock URL
  \url{https://www.kaggle.com/c/walmart-recruiting-store-sales-forecasting}.

\bibitem[Abadie and Gardeazabal(2003)]{abadie3}
A.~Abadie and J.~Gardeazabal.
\newblock The economic costs of conflict: A case study of the basque country.
\newblock \emph{American Economic Review}, 2003.

\bibitem[Abadie et~al.(2010)Abadie, Diamond, and Hainmueller]{abadie1}
A.~Abadie, A.~Diamond, and J.~Hainmueller.
\newblock Synthetic control methods for comparative case studies: Estimating
  the effect of california{\^a}s tobacco control program.
\newblock \emph{Journal of the American Statistical Association}, 2010.

\bibitem[Abadie et~al.(2011)Abadie, Diamond, and Hainmueller]{abadie2}
A.~Abadie, A.~Diamond, and J.~Hainmueller.
\newblock Synth: An r package for synthetic control methods in comparative case
  studies.
\newblock \emph{Journal of Statistical Software}, 2011.

\bibitem[Agarwal et~al.(2018)Agarwal, Amjad, Shah, and
  Shen]{timeseriesMatrixEst}
Anish Agarwal, Muhammad~Jehangir Amjad, Devavrat Shah, and Dennis Shen.
\newblock Model agnostic time series analysis via matrix estimation.
\newblock \emph{Proc. ACM Meas. Anal. Comput. Syst.}, 2\penalty0 (3):\penalty0
  40:1--40:39, December 2018.
\newblock URL \url{http://doi.acm.org/10.1145/3287319}.

\bibitem[Agarwal et~al.(2019)Agarwal, Shah, Shen, and Song]{asss}
Anish Agarwal, Devavrat Shah, Dennis Shen, and Dogyoon Song.
\newblock On robustness of principal component regression.
\newblock \emph{arXiv preprint}, 2019.

\bibitem[Amjad et~al.(2018)Amjad, Shah, and Shen]{rsc1}
Muhammad Amjad, Devavrat Shah, and Dennis Shen.
\newblock Robust synthetic control.
\newblock \emph{Journal of Machine Learning Research}, 19\penalty0
  (17):\penalty0 1--50, 2018.

\bibitem[Amjad(2018)]{amjad2018}
Muhammad~J Amjad.
\newblock \emph{Sequential Data Inference via Matrix Estimation: Causal
  Inference, Cricket and Retail}.
\newblock PhD thesis, Massachusetts Institute of Technology, 2018.

\bibitem[Bailey and Clarke(2006)]{Bailey2006}
Michael Bailey and Stephen~R Clarke.
\newblock Predicting the match outcome in one day international cricket
  matches, while the game is in progress.
\newblock \emph{Journal of Sports Science \& Medicine}, 5\penalty0
  (4):\penalty0 480--487, 12 2006.
\newblock URL \url{http://www.ncbi.nlm.nih.gov/pmc/articles/PMC3861745/}.

\bibitem[Bareinboim and Pearl(2016)]{Bareinboim7345}
Elias Bareinboim and Judea Pearl.
\newblock Causal inference and the data-fusion problem.
\newblock \emph{Proceedings of the National Academy of Sciences}, 113\penalty0
  (27):\penalty0 7345--7352, 2016.
\newblock ISSN 0027-8424.
\newblock \doi{10.1073/pnas.1510507113}.
\newblock URL \url{https://www.pnas.org/content/113/27/7345}.

\bibitem[Borgs et~al.(2017)Borgs, Chayes, Lee, and Shah]{BorgsChayesLeeShah17}
Christian Borgs, Jennifer Chayes, Christina~E Lee, and Devavrat Shah.
\newblock Thy friend is my friend: Iterative collaborative filtering for sparse
  matrix estimation.
\newblock In \emph{Advances in Neural Information Processing Systems}, pages
  4718--4729, 2017.

\bibitem[Chatterjee(2015)]{usvt}
Sourav Chatterjee.
\newblock Matrix estimation by universal singular value thresholding.
\newblock \emph{Annals of Statistics}, 43:\penalty0 177--214, 2015.

\bibitem[Chen and Cichocki(2005)]{chen2005}
Zhe Chen and Andrzej Cichocki.
\newblock Nonnegative matrix factorization with temporal smoothness and/or
  spatial decorrelation constraints.
\newblock \emph{Laboratory for Advanced Brain Signal Processing, RIKEN, Tech.
  Rep}, 68, 2005.

\bibitem[Efron and Morris(1977)]{stein1}
Bradley Efron and Carl Morris.
\newblock Stein's paradox in statistics.
\newblock 236:\penalty0 119--127, 05 1977.

\bibitem[Hogan(2015)]{wasp1}
Seamus Hogan.
\newblock Cricket and the wasp: Shameless self promotion (wonkish)., January
  2015.
\newblock URL
  \url{http://offsettingbehaviour.blogspot.com/2012/11/cricket-and-wasp-shameless-self.html}.

\bibitem[Lee(2017)]{lee2017}
Christina Lee.
\newblock \emph{Latent Variable Model Estimation via Collaborative Filtering}.
\newblock PhD thesis, Massachusetts Institute of Technology, 2017.

\bibitem[Oliver et~al.(2018)Oliver, Basevi, Luke, and Wilde]{cricviz}
Phil Oliver, Travis Basevi, Will Luke, and Freddie Wilde.
\newblock Cricviz, 2018.
\newblock URL \url{http://cricviz.com/}.

\bibitem[Paul R.~Rosenbaum(1983)]{rosenbaum1983}
Donald B.~Rubin Paul R.~Rosenbaum.
\newblock The central role of the propensity score in observational studies for
  causal effects.
\newblock \emph{Biometrika}, 70\penalty0 (1):\penalty0 41--55, 1983.

\bibitem[Pearl(2009)]{pearl2009}
Judea Pearl.
\newblock Causal inference in statistics: An overview.
\newblock \emph{Statistics Surveys}, 3:\penalty0 96--146, 2009.

\bibitem[Rallapalli et~al.(2010)Rallapalli, Qiu, Zhang, and
  Chen]{rallapalli2010}
Swati Rallapalli, Lili Qiu, Yin Zhang, and Yi-Chao Chen.
\newblock Exploiting temporal stability and low-rank structure for localization
  in mobile networks.
\newblock In \emph{Proceedings of the Sixteenth Annual International Conference
  on Mobile Computing and Networking}, MobiCom '10, pages 161--172, New York,
  NY, USA, 2010. ACM.
\newblock ISBN 978-1-4503-0181-7.
\newblock \doi{10.1145/1859995.1860015}.
\newblock URL \url{http://doi.acm.org/10.1145/1859995.1860015}.

\bibitem[Rubin(1973)]{rubin1973}
Donald~B. Rubin.
\newblock Matching to remove bias in observational studies.
\newblock \emph{Biometrics}, 29\penalty0 (1):\penalty0 159--183, 1973.

\bibitem[Rubin(1974)]{rubin1974}
Donald~B. Rubin.
\newblock Estimating causal effects of treatments in randomized and
  nonrandomized studies.
\newblock \emph{Journal of Educational Psychology}, 66\penalty0 (5):\penalty0
  688--701, 1974.

\bibitem[Shemilt(2015)]{bbccricket}
Stephan Shemilt.
\newblock Cricket world cup 2015: India and pakistan fans usurp the limelight,
  February 2015.
\newblock URL \url{https://www.bbc.com/sport/cricket/31479487}.

\bibitem[Tharoor(2016)]{wapocricket}
Ishan Tharoor.
\newblock These global sporting events totally dwarf the superbowl, February
  2016.
\newblock URL
  \url{https://www.washingtonpost.com/news/worldviews/wp/2016/02/05/these-global-sporting-events-totally-dwarf-the-super-bowl/?noredirect=on&utm_term=.6f393cc4e52e}.

\bibitem[Xie et~al.(2016)Xie, Talk, and Fox]{xie16missingtimeseries}
Christopher Xie, Alex Talk, and Emily Fox.
\newblock A unified framework for missing data and cold start prediction for
  time series data.
\newblock In \emph{Advances in neural information processing systems Time
  Series Workshop}, 2016.

\bibitem[Yu et~al.(2015)Yu, Rao, and Dhillon]{dhillon2016}
Hsiang{-}Fu Yu, Nikhil Rao, and Inderjit~S. Dhillon.
\newblock Temporal regularized matrix factorization.
\newblock \emph{CoRR}, abs/1509.08333, 2015.

\end{thebibliography}

\newpage

\appendix
\section{Proofs} \label{sec:proofs}

\subsection{Proof of Proposition \ref{prop:low_rank_approx}}

\begin{proof}
	We will construct a low-rank tensor $\tT$ by partitioning the latent row and tube spaces of $\tM$. Through this process, we will demonstrate that each frontal slice of $\tT$ is a low-rank matrix, and only a subset of the $K$ frontal slices of $\tT$ are distinct. Together, these observations establish the low-property of $\tT$. Finally, we will complete the proof by showing that $\tM$ is entry-wise arbitrarily close to $\tT$.  \\
	
{\bf Partitioning the latent space to construct $\tT$.} 
Fix some $\delta_1, \delta_3 > 0$. Since the latent row parameters $\theta_i$ come from a compact space $[0,1]^{d_1}$, we can construct a finite covering (partition) $P(\delta_1) \subset [0,1]^{d_1}$ such that for any $\theta_i \in [0,1]^{d_1}$, there exists a $\theta_{i'} \in P(\delta_1)$ satisfying $\norm{\theta_i - \theta_{i'}}_2 \le \delta_1$. By the same argument, we can construct a partitioning $P(\delta_3) \subset [0,1]^{d_3}$ such that $\norm{\omega_k - \omega_{k'}}_2 \le \delta_3$ for any $\omega_k \in [0,1]^{d_3}$ and some $\omega_{k'} \in P(\delta_3)$. By the Lipschitz property of $f$ and the compactness of the latent space, it follows that $\abs{P(\delta_1)} \le C_1 \cdot \delta_1^{-d_1}$, where $C_1$ is a constant that depends only on the space $[0,1]^{d_1}$, ambient dimension $d_1$, and Lipschitz constant $\mathcal{L}$. Similarly, $\abs{P(\delta_3)} \le C_3 \cdot \delta_3^{-d_3}$, where $C_3$ is a constant that depends only on $[0,1]^{d_3}$, $d_3$, and $\mathcal{L}$. \\

For each $\theta_i$, let $p_1(\theta_i)$ denote the unique element in $P(\delta_1)$ that is closest to $\theta_i$. At the same time, we define $p_3(\omega_k)$ as corresponding element in $P(\delta_3)$ that is closest to $\omega_k$. We now construct our tensor $\tT = [T_{ijk}]$ by defining its $(i,j,k)$-th entry as 
\begin{align*}
	T_{ijk} =  f(p_1(\theta_i), \rho_j, p_3(\omega_k))
\end{align*}
for all $i\in [N]$, $j \in [T]$, and $k \in [K]$. \\

{\bf Establishing the low-rank property of $\tT$.} 
Let us fix a frontal slice $k \in [K]$. Consider any two rows of $\tT_{\cdot, \cdot, k}$, say $i$ and $i'$. If $p_1(\theta_i) = p_1(\theta_{i'})$, then rows $i$ and $i'$ of $\tT_{\cdot, \cdot, k}$ are identical. Hence, there at most $\abs{P(\delta_1)}$ distinct rows in $\tT_{\cdot, \cdot, k}$, i.e., $\text{rank}(\tT_{\cdot, \cdot, k}) \le \abs{P(\delta_1)}$. In words, each frontal slice of $\tT$ is a low-rank matrix with its rank bounded above by $\abs{P(\delta_1)}$. \\

Now, consider any two frontal slices $k$ and $k'$ of $\tT$. If $p_3(\omega_k) = p_3(\omega_{k'})$, then for all $i \in [N]$ and $j \in [T]$, we have 
\begin{align*}
	T_{ijk} &= f(p_1(\theta_i), \rho_j, p_3(\omega_k)) = f(p_1(\theta_i), \rho_j, p_3(\omega_{k'})) = T_{ijk'}.
\end{align*}
In words, the $k$-th frontal slice of $\tT$ is equivalent to the $k'$-th frontal slice of $\tT$. Hence, $\tT$ has at most $\abs{P(\delta_3)}$ distinct frontal slices. \\

To recap, we have established that all of the frontal slices $\tT_{\cdot, \cdot, k}$ of $\tT$ are low-rank matrices, and only a subset of the frontal slices of $\tT$ are distinct. Therefore, it follows that the rank of $\tT$ (i.e., the smallest integer $r$ such that $\tT$ can be expressed as a sum of rank one tensors), is bounded by the product of the maximum matrix rank of any slice $\tT_{\cdot, \cdot, k}$ of $\tT$ with the number of distinct slices in $\tT$. More specifically, if we let $\delta = \delta_1 = \delta_3$, then
\begin{align*}
	\text{rank}(\tT) &\le \abs{P(\delta_1)} \cdot \abs{P(\delta_3)} \le C \cdot \delta^{-(d_1 + d_3)},
\end{align*}
where $C$ is a constant that depends on the latent spaces $[0,1]^{d_1}$ and $[0,1]^{d_3}$, the dimensions $d_1$ and $d_3$, and the Lipschitz constant $\mathcal{L}$. We highlight that the bound on the tensor rank does not depend on the dimensions of $\tT$. \\

{\bf $\tM$ is well approximated by $\tT$.}
Here, we bound the maximum difference of any entry in $\tM = [M_{ijk}]$ from $\tT = [T_{ijk}]$. Using the Lipschitz property of $f$, for any $(i,j,k) \in [N] \times [T] \times [K]$, we obtain
\begin{align*}
	\abs{M_{ijk} - T_{ijk}} &= \abs{ f(\theta_i, \rho_j, \omega_k) - f(p_1(\theta_i), \rho_j, p_3(\omega_k))} 
	\\ &\le \mathcal{L} \cdot \left(\norm{\theta_i - p_1(\theta_i)}_2 + \norm{\omega_k - p_3(\omega_k)}_2 \right) 
	\\ &\le \mathcal{L} \cdot (\delta_1 + \delta_3). 
\end{align*}
This establishes that $\tM$ is entry-wise arbitrarily close to $\tT$. Setting $\delta = \delta_1 = \delta_3$ completes the proof. 

\end{proof}

\subsection{Proof of Proposition \ref{prop:linear_comb}}

\begin{proof}
Recall the definition of $\bU$ from \eqref{eq:low_rank_tensor}. By Property \ref{property:low_rank}, 
we have that $\text{dim}(\text{span}\{ \bU_{1, \cdot}, \dots, \bU_{N, \cdot} \}) = r$, where $\bU_{i, \cdot}$ 
denotes the $i$-th row of $\bU$. Since we are choosing the treatment unit uniformly at random amongst 
the $N$ possible indices (due to the re-indexing of indices as per some random permutation), the probability of that $\bU_{1, \cdot}$ is not a linear combination 
of the other rows is $r / N$. In light of this observation, we define 
$A = \{ \bU_{1, \cdot} = \sum_{z=2}^N \beta^*_z \bU_{z, \cdot} \}$ as the event where the first 
row of $\bU$ is a linear combination of the other rows in $\bU$. Then, using the arguments above, 
we have that $\Pb \{A \} = 1 - \frac{r}{N}$. \\
	
Now, suppose the event $A$ occurs. Then for all $j \in [T]$ and $k \in [K]$,
	\begin{align*}
		M_{1jk} &= \sum_{i=1}^r U_{1i} \cdot V_{ji} \cdot W_{ki}
		\\ &= \sum_{i=1}^r \left( \sum_{z=2}^N \beta^*_z \cdot U_{zi} \right) \cdot V_{ji} \cdot W_{ki}
		\\ &= \sum_{z=2}^N \beta^*_z \cdot  \left( \sum_{i=1}^r U_{zi} \cdot V_{ji} \cdot W_{ki} \right)
		\\ &= \sum_{z=2}^N \beta^*_z \cdot M_{zjk}.
	\end{align*}
	This completes the proof. 
\end{proof}

\subsection{Proof of Theorem \ref{thm:pre_int}} 

\begin{proof}
The proof follows from an immediate application of Theorem 3 of \cite{asss}. 
\end{proof}


\subsection{Proof of Theorem \ref{thm:post_int}} 

\begin{proof}
	The proof follows from an immediate application of Theorem 5 of \cite{asss}. 
\end{proof}

\end{document}